\documentstyle[12pt]{article}
\renewcommand{\theequation}{\arabic{equation}}
\newcommand{\EQ}{\begin{equation}}
\newcommand{\EN}{\end{equation}}

\newcommand{\ket}[1]{\left|#1\right\rangle}      

\newcommand{\bear}{\begin{eqnarray}}
\newcommand{\ear}{\end{eqnarray}}

\begin{document}

\topmargin 0pt
\oddsidemargin 5mm
\newcommand{\NP}[1]{Nucl.\ Phys.\ {\bf #1}}
\newcommand{\PL}[1]{Phys.\ Lett.\ {\bf #1}}
\newcommand{\NC}[1]{Nuovo Cimento {\bf #1}}
\newcommand{\CMP}[1]{Comm.\ Math.\ Phys.\ {\bf #1}}
\newcommand{\PR}[1]{Phys.\ Rev.\ {\bf #1}}
\newcommand{\PRL}[1]{Phys.\ Rev.\ Lett.\ {\bf #1}}
\newcommand{\MPL}[1]{Mod.\ Phys.\ Lett.\ {\bf #1}}
\newcommand{\JETP}[1]{Sov.\ Phys.\ JETP {\bf #1}}
\newcommand{\TMP}[1]{Teor.\ Mat.\ Fiz.\ {\bf #1}}
     
\renewcommand{\thefootnote}{\fnsymbol{footnote}}
     
\newpage
\setcounter{page}{0}
\begin{titlepage}     
\begin{flushright}
UFSCARF-TH-96-03
\end{flushright}
\vspace{0.5cm}
\begin{center}
{\large One parameter family of an integrable $spl(2|1)$ vertex model: 
Algebraic Bethe ansatz and ground state structure}\\
\vspace{1cm}
\vspace{1cm}
{\large  P.B. Ramos and M.J.  Martins } \\
\vspace{1cm}
{\em Universidade Federal de S\~ao Carlos\\
Departamento de F\'isica \\
C.P. 676, 13560~~S\~ao Carlos, Brasil}\\
\end{center}
\vspace{1.2cm}
     
\begin{abstract}
We formulate in terms of the quantum inverse scattering method the exact 
solution of a $spl(2|1)$ invariant vertex model recently introduced in the
literature. The corresponding transfer matrix is diagonalized by using the
algebraic (nested) Bethe ansatz approach. The ground state structure is 
investigated and we argue that a   
Pokrovsky-Talapov 
transition is favored for certain value of the 4-dimensional $spl(2|1)$ 
parameter. 
\end{abstract}
\vspace{.2cm}
\vspace{.2cm}
\centerline{February 1996}
\end{titlepage}

\renewcommand{\thefootnote}{\arabic{footnote}}
\section{Introduction}

Recently a novel solution of the Yang-Baxter equation invariant under
the $spl(2|1)$ symmetry has been obtained in the literature by Bracken et al
and Maassarani \cite{BA,MA,BA1}. Its peculiar feature is the presence of an
additional non-additive parameter whose origin goes back to the continuous 
$4$-dimensional irreducible representation of the algebra  $spl(2|1)$ \cite{RI} \footnote{We remark that the $R$-matrix in the context of the fundamental 
representation of $OSP(2|2)$ algebra has been previously discussed  in ref. \cite{ITO}}.
The physical meaning of this solution is that the associated quantum one-dimensional
Hamiltonian \cite{BA1} can be interpreted as an exactly solved model of strongly
correlated electrons possessing extra fine-tuned hopping terms and electron
pair interactions besides the typical correlations appearing in the Hubbard 
model. Most of the Bethe ansatz results concerning this model were obtained either by using the coordinate Bethe ansatz method directly on the fermionic
Hamiltonian \cite{FRA,KU} or by taking advantage of the fusion construction 
of the $4$-dimensional $spl(2|1)$ $R$-
matrix. 
\cite{MA}. In this last case, we note that the eigenvalues of the corresponding
transfer matrix and the Bethe ansatz equations have appeared only as reasonable
conjectures. We also remark that the authors \cite{MP} managed to present the
phenomenological analytical Bethe ansatz approach for a special point of the
continuous parameter in which the $R$-matrix can be seen as a braid-monoid 
invariant. However, for this interesting system, a more unified approach such as the quantum inverse scattering framework \cite{FAD,QI,DV} has not yet 
succeeded to be formulated. We 
remark that the quantum inverse scattering approach 
makes it possible to present the Bethe ansatz results in an elegant way: the
Bethe ansatz equations and the eigenvalues of the transfer matrix appear as a 
consequence of systematic algebraic manipulations of the creation and 
annihilation operators acting on a certain pseudovacuum. The main purpose of 
this paper is to fill this gap by presenting a detailed formulation of the
quantum inverse scattering method for such new $R$-matrix invariant by the
superalgebra $spl(2|1)$. For sake of simplicity, we are going to 
describe our formulation for the rational limit of the $spl(2|1)$ 
$R$-matrix. Some numerical and analytical results concerning the
 ground state structure of this model are also presented.

	This paper is organized as follow. The next section is concerned with 
the presentation of the Boltzmann weights, the associated Hamiltonian and the
basic properties of the $spl(2|1)$ rational $R$-matrix. In section $3$ we
formulate the quantum inverse scattering method and 
in section $4$ we obtain several important
commutation relations. In section $5$ we elaborate on the
construction of the eigenstates and it is shown that the eigenvalues
depend on an extra auxiliary problem of diagonalization. The corresponding
nested Bethe ansatz equations and the eigenvalues are then 
determined in section $6$.  
In section $7$ we
discuss the ground state structure of this model and we present some evidence
that a Pokrovsky-Talapov transition may occur for a special value of the
non-additive parameter of the $spl(2|1)$ $R$-matrix. Section $8$ is reserved
for our conclusions and remarks concerning the application of our formulation
to other integrable systems. In 
appendices $A$, $B$ and $C$ we have collected some useful relations concerning the $spl(2|1)$ algebra, the two-particle state, and the three-particle state, 
respectively .

\section{The $spl(2|1)$ vertex model and its properties}

We start this section by describing the Boltzmann weights of the 
$spl(2|1)$ $R$-matrix recently proposed in refs. \cite{BA,MA,BA1}. We shall 
consider the rational limit of the $spl(2|1)$ $R$-matrix, avoiding certain
extra mathematical manipulations typical of trigonometric weights. We note,
however, that the vertex structure is basically the same for both rational and
trigonometrical cases and therefore many of the ideas described in this paper
are quite general. Here we will adopt the notations of the appendix of ref.
\cite{MA}. The rational $spl(2|1)$ $R$-matrix can be written in terms of 
certain combination of projectors $P_{1}(b)$ and $P_{3}(b)$ \cite{MA} of the
$spl(2|1)$ algebra as follows:
\EQ
R(\lambda,b) = I - \frac{4\lambda}{1-2b+2\lambda} P_{1}(b) -\frac{4\lambda}{1+2b+2\lambda} P_{3}(b)
\EN
where $\lambda$ is the spectral parameter, b characterizes the continuous 
$4$-dimensional representation of the $spl(2|1)$ algebra \footnote{We remark
that at points $b= \pm 1/2$ the corresponding one-dimensional Hamiltonian is 
singular. Here we exclude of our analysis such atypical representation.} and 
I is the identity operator. For completeness we present the explicit matrix
expression of the projectors $P_{1}(b)$ e $P_{3}(b)$ in Appendix A. Using these
formulas one can see that the 
rational $spl(2|1)$ $R$-matrix consists of $36$ 
non-null Boltzmann weights. We have schematized them in figure $1$. We choose to
represent each bond of the lattice with the variables $f_{i}$ and $b_{i}$
($i=1,2$) in order to represent the fermionic $(f_{1},f_{2})$ and the bosonic
$(b_{1},b_{2})$ degrees of freedom used by Maassarani \cite{MA}. More precisely,
the $spl(2|1)$ $R$-matrix can be considered as a matrix acting on the tensor
product of the two $4$-dimensional auxiliary space 
$C^{4} \times C^{4}$ and can be
arranged as a $16 \times 16$ matrix which in the $fbbf$ 
grading possesses the following
form 

\EQ
{\footnotesize
R(\lambda,b) =
  \pmatrix{
 l(\lambda)&0 &0 &0 &0  &0  &0  &0  &0  &0  &0  &0  &0  &0  &0  &0    \cr
 0 & m(\lambda) &0 &0 &f(\lambda)  &0  &0  &0  &0  &0  &0  &0  &0  &0  &0  &0    \cr
 0 &0 &m(\lambda) &0 &0  &0  &0  &0  &f(\lambda)  &0  &0  &0  &0  &0  &0  &0    \cr
 0 &0 &0 &n(\lambda) &0  &0  &-\sigma(\lambda)  &0  &0  &\sigma(\lambda)  &0  &0  &p(\lambda)  &0  &0  &0    \cr
 0 &f(\lambda) &0 &0 &m(\lambda)  &0  &0  &0  &0  &0  &0  &0  &0  &0  &0  &0    \cr
 0 &0 &0 &0 &0  &1  &0  &0  &0  &0  &0  &0  &0  &0  &0  &0    \cr
 0 &0 &0 &\sigma(\lambda) &0  &0  &s(\lambda)  &0  &0  &t(\lambda)  
&0  &0  &\sigma(\lambda)  &0  &0  &0    \cr
 0 &0 &0 &0 &0  &0  &0  &q(\lambda)  &0  &0  &0  &0  &0  &g(\lambda)  &0  &0    \cr
 0 &0 &f(\lambda) &0 &0  &0  &0  &0  &m(\lambda)  &0  &0  &0  &0  &0  &0  &0    \cr
 0 &0 &0 &-\sigma(\lambda) &0  &0  &t(\lambda)  &0  &0  &s(\lambda)  &0  &0  &-\sigma(\lambda)  &0  &0  &0    \cr
 0 &0 &0 &0 &0  &0  &0  &0  &0  &0  &1  &0  &0  &0  &0  &0    \cr
 0 &0 &0 &0 &0  &0  &0  &0  &0  &0  &0  &q(\lambda)  &0  &0  &g(\lambda)  &0    \cr
 0 &0 &0 &p(\lambda) &0  &0  &-\sigma(\lambda)  &0  &0  &\sigma(\lambda)  &0  &0  &n(\lambda)  &0  &0  &0    \cr
 0 &0 &0 &0 &0  &0  &0  &g(\lambda)  &0  &0  &0  &0  &0  &q(\lambda)  &0  &0    \cr
 0 &0 &0 &0 &0  &0  &0  &0  &0  &0  &0  &g(\lambda)  &0  &0  &q(\lambda)  &0    \cr
 0 &0 &0 &0 &0  &0  &0  &0  &0  &0  &0  &0  &0  &0  &0  &r(\lambda)    \cr}
}
\EN
In the expression $(2)$ the functional dependence of the Boltzmann weights is 
given by
\begin{displaymath}
l(\lambda) = {1-2b-2\lambda \over 1-2b+2\lambda},~~
r(\lambda) = {1+2b-2\lambda \over 1+2b+2\lambda},~~
n(\lambda) = {(1-2b)(1+2b) \over (1-2b+2\lambda)(1+2b+2\lambda)} 
\end{displaymath}
\begin{displaymath}
m(\lambda) = {1-2b \over 1-2b+2\lambda},~~
q(\lambda) = {1+2b \over 1+2b+2\lambda},~~
p(\lambda) = {-4\lambda(\lambda+1) \over (1-2b+2\lambda)
(1+2b+2\lambda)} 
\end{displaymath}
\begin{displaymath}
f(\lambda) = {2\lambda \over 1-2b+2\lambda},~~
g(\lambda) = {2\lambda \over 1+2b+2\lambda},~~
t(\lambda) = {4\lambda^{2} \over (1+2b+2\lambda)(1-2b+2\lambda)} 
\end{displaymath}
\EQ
\sigma(\lambda) = 
{2\lambda[(1-2b)(1+2b)]^{1/2} \over (1+2b+2\lambda)(1-2b+2\lambda)},~~
s(\lambda) = {1-4b^{2}+4\lambda \over (1+2b+2\lambda)(1-2b+2\lambda)}
\EN
%

	A general feature of the $R$-matrix $(2)$ is that many of their 
Boltzmann weights are not invariant under the charge $1 \leftrightarrow 2$ 
symmetry for same specie of bosonic and fermionic index, namely $f_1 \leftrightarrow f_2$ and $b_1 \leftrightarrow b_2$. For instance, from figure
$1$ and expressions $(3)$, these are the cases of the following pairs of 
weights $\{l(\lambda),r(\lambda)\}$,$\{ 
m(\lambda),q(\lambda)\}$ and $\{f(\lambda),g(\lambda)\}$.
In such case, a standard crossing symmetry cannot be implemented, since 
it is not possible to find common crossing factor for the weights $l(\lambda)$
and $r(\lambda)$, for example. In fact, under the  $1 \leftrightarrow 2$ symmetry
 the parameter b is reflected to $-b$. Hence, the remaining invariance is the reflection $b \rightarrow -b$, and the physical properties derived either of 
$R(\lambda,b)$ or of $R(\lambda,-b)$ 
should  in fact be the same. We remark, however, that the point $b=0$ is
clearly an exception 
of the discussion we have 
made above. For this special point, the $1 \leftrightarrow 2$ symmetry
is present, usual crossing relations are then possible to be established and consequently a 
interpretation of $R(\lambda,b=0)$ in terms of a factorizable $S$-matrix turns out to be possible 
\cite{MP}.

Further properties of the $R$-matrix $(2)$ can 
be seen by using certain relations satisfied by the
projectors $P_{1}(b)$ and $P_{3}(b)$. We have collected such important 
relations in Appendix $A$. For example, 
one can show that the quasi-classical $r$-matrix 
originated from the $R$-matrix $(1)$ encodes the $spl(2|1)$ symmetry in a
standard way, namely in terms of the $spl(2|1)$ Casimir operator. In fact, making the redefinition
$\lambda \rightarrow \frac{\lambda}{\eta}$ and expanding 
around  $\eta =0$ we find
\EQ
R(\lambda,b,\eta) \sim P^{g} \left [ 1 + \frac{\eta (C(b) - I)}{2\lambda}
\right ]
\EN
where $\eta$ is the classical parameter, $C(b)$ is the Casimir operator of $spl(2|1)$ and $P^{g}$
is the graded permutation operator. Details of this 
calculation can be found in Appendix $A$. Another
interesting feature of such $R$-matrix is its braid-monoid representation at the special point $b=0$.
In particular, a Temperley-Lieb operator $E$ $(monoid)$ can be defined in terms of the projectors 
$P_{1}(b)$ and $P_{3}(b)$ as follows
\EQ
E = \lim_{b \rightarrow 0} 4b[P_{1}(b) - P_{3}(b)]
\EN
which together with the permutation 
operator $P^{g}$ satisfy the braid-monoid relation \cite{MP}. In 
this sense, it seems  very interesting to find the underlying 
algebraic structure for the points $b \neq 0$. After Baxterization such
general structure should produce the $spl(2|1)$ $R$-matrix (2). 
A precise answer to this question
has eluded us so far.

\section{The quantum inverse scattering approach}

The main purpose of this section is to begin the formulation of the eigenvalue problem of the
corresponding transfer matrix $T(\lambda)$ of the 
$spl(2|1)$ vertex system on a square lattice of
size $L \times L$. The diagonalization problem
\EQ
T(\lambda) \ket{\Phi} = \Lambda(\lambda) \ket{\Phi}
\EN
can be solved by using an algebraic construction \cite{FAD,QI,DV} based on the Yang-Baxter algebra of 
monodromy matrices ${\cal T}(\lambda)$
\EQ
R(\lambda - \mu) {\cal T}(\lambda) \otimes {\cal T}(\mu) =
{\cal T}(\mu) \otimes {\cal T}(\lambda) R(\lambda - \mu)
\EN
where the matrix ${\cal T}(\lambda)$ acts on the tensor product of an auxiliary space and a quantum 
space $C^{4} \otimes C^{4L}$ and is given in terms of the product of vertex operators 
${\cal L}(\lambda)$ by
\EQ
{\cal T}(\lambda) = {\cal L}_{oL}(\lambda){\cal L}_{oL-1}(\lambda)....{\cal L}_{o1}(\lambda)
\EN
where the index `$o$' stands for the $4 \times 4$ auxiliary space, and as usual the transfer matrix $T(\lambda)$
is obtained as a trace of the monodromy matrix ${\cal T}(\lambda)$ over such auxiliary space. The
elements of the vertex operator $L_{ab}^{cd}(\lambda)$ are related to those of the $spl(2|1)$ $R$-
matrix (2) by a permutation on the $C^{4} \times C^{4}$ tensor space
\EQ
L_{ab}^{cd}(\lambda) = R_{ba}^{cd}(\lambda)
\EN
	The corresponding quantum one-dimensional Hamiltonian can be obtained as the logarithmic
derivative of the transfer matrix $T(\lambda)$ at the regular point $\lambda = 0$. After some
algebraic manipulation (see Appendix A), the associated spin 
chain can be only written  in terms
of the Casimir operator of the $spl(2|1)$ algebra by the following expression
\EQ
H = -\frac{2}{(2b+1)^2 (1-2b)^2}\sum_{i=1}^{L} [ 2(1+2b^2)I - 
(1+4b^{2})C_{i,i+1}(b) 
- C_{i,i+1}^2 (b) ]
\EN
where we assumed standard periodic boundary conditions. As a 
consequence, we can see that the reflection symmetry $ b \rightarrow -b $ is explicitly exhibited
by expression (10) since there exists an isomorphism between 
$ C_{i,i+1}(b) $ and $ C_{i,i+1}(-b) $ ( see end of appendix $A$ ) .

Before going on, we remark that the $R$-matrix (2) is a null-parity (Grassmann) braid operator,
and after some sign definitions\footnote{The graded $\hat{R}$-matrix
is related to the standard one (2) by $\hat{R}_{ba}^{cd}(\lambda) =
(-1)^{p(a)p(b)}R_{ab}^{cd}(\lambda), $ where $p(i)$ denotes the Grasmmann 
parity of $i=1, \cdots,4$.} 
produces a vertex operator which solves the graded Yang-Baxter 
equation \cite{KR}. In this case one has to use the supersymmetric 
formalism developed in refs. \cite{KR} by
basically changing standard properties 
such as trace and tensor product by their analogs on the 
graded spaces. However, a graded formulation does not simplify 
the original problem of diagonalization
of the transfer matrix. On 
the contrary, in terms of practical calculations with the corresponding 
Hamiltonian, one has to be very careful 
to keep track of the fermions signs appearing on the tensor product
of the Hilbert space. Here we would like to stress that the 
diagonalization of the corresponding
Hamiltonian for small lattice sizes is extremely 
important as a guideline, giving us a correct  
insight of basic properties such as the ground 
state structure. For that reason we stick with the
standard formalism, in which such job 
can be performed in a more direct and safe way. Anyhow, the 
basic difference between the standard and graded formulation will be the
presence of extra phase factors on the Bethe ansatz equations. However, such phase-factors can be
accomplished as general boundary 
conditions similarly as has been done before for the  graded $OSP(1|2)$ model
\cite{OSP}.

After presenting the basic definitions and discussions concerning 
the diagonalization problem
(6), we are going to turn our attention to the construction of the eigenstates $\ket{\Phi}$ and eigenvalues 
$\Lambda(\lambda)$, respectively. We start our discussion by solving 
the commutation relations which follows as a consequence of
the Yang-Baxter algebra.

\section{The fundamental commutation relations}

The proper way to work out the 
intertwining relations (7) depends, to some extent,
on the properties of the vertex operator ${\cal L}(\lambda)$ when it 
acts on a given reference state. Let
us  consider as the reference state the usual ferromagnetic pseudovacuum given by
\EQ
\ket{0} = \prod_{i=1}^{L} \otimes \ket{0}_{i} , ~~
\ket{0}_{i} =
\pmatrix{
1 \cr
0 \cr
0 \cr
0 \cr}
\EN

By using equations (2) and (9) we find that the 
vertex operator ${\cal L}(\lambda)$ acting on 
the state $\ket{0}_{i}$ has the important triangular form
\EQ 
{\cal L}(\lambda)\ket{0}_{i} =
\pmatrix{
l(\lambda)  &  *  &  *  &  *  \cr
0  &  f(\lambda)  &  0  &  *  \cr
0  &  0  &  f(\lambda)  &  *  \cr
0  &  0  &  0  &  p(\lambda)  \cr}
\ket{0}_i
\EN

Now, if we write the monodromy matrix ${\cal T}(\lambda)$ as a $4 \times 4$ 
matrix having the 
following particular form
\EQ
{\cal T}(\lambda) =
\pmatrix{
B(\lambda)   &   B_{1}(\lambda)   &   B_{2}(\lambda)   &   F(\lambda)   \cr
C_{1}(\lambda)  &  A_{11}(\lambda)  &  A_{12}(\lambda)  &  E_{1}(\lambda)   \cr
C_{2}(\lambda)  &  A_{21}(\lambda)  &  A_{22}(\lambda)  &  E_{2}(\lambda)   \cr
C_{3}(\lambda)  & C_{4}(\lambda)  &  C_{5}(\lambda)  &  D(\lambda)  \cr}
\EN
the problem of diagonalization of the transfer matrix becomes
\EQ
[B(\lambda)+\sum_{1}^{2}A_{aa}(\lambda)+D(\lambda)] \ket{\Phi} = 
\Lambda(\lambda) \ket{\Phi}
\EN

Moreover, as a consequence of definition (8),  we find that the 
following diagonal relations are also satisfied
\EQ
B(\lambda)\ket{0} = [l(\lambda)]^{L}\ket{0},~~ D(\lambda)\ket{0} = 
p(\lambda)^{L}\ket{0},~~ A_{aa}(\lambda)\ket{0} = f(\lambda)^{L}\ket{0} , a=1,2
\EN
as well as the annihilation properties
\EQ
C_{i}(\lambda)\ket{0} = 0~(i=1,\cdots,5),~~ A_{ab}(\lambda)\ket{0} = 0~(a \neq b =1,2)
\EN

In particular for the eigenstate $\ket{0}$ the eigenvalue $\Lambda(\lambda)$ is determined to be
\EQ
\Lambda(\lambda) = [l(\lambda)]^{L} + \sum_{a=1}^{2}[f(\lambda)]^{L} + 
[p(\lambda)]^{L}
\EN

In order to construct other eigenvalues one has to find the commutation rules between the 
operators appearing in definition (13). Unlikely to what happens 
to the $6$-vertex model \cite{FAD,QI,DV} and
its multi-state generalizations \cite{DV1,RU}, some of the `nice' 
commutation relations of the
$spl(2|1)$ vertex model are more complicated and require  additional
work. In many 
cases one needs to combine in a special way certain relations in order to
get the appropriate commutation rule. Let us illustrate the main idea for the particular case of the commutation relations 
between the operators $A_{ab}(\lambda)$ and $B_{c}(\lambda)$. As usual, we 
substitute the form of the
monodromy matrix (13) in the intertwining equation (7) and by 
using the Boltzmann weights of the 
$R$-matrix (2) we find the relation
\bear
A_{ab}(\lambda)B_{c}(\mu) = \frac{1}{f(\lambda-\mu)}\tilde{r}_{ed}^{bc}(\lambda-u)B_{e}(\mu)A_{ad}(\lambda) 
- \frac{m(\lambda-\mu)}{f(\lambda-\mu)} B_{b}(\lambda)A_{ac}(\mu) 
 \nonumber \\
- \xi_{bc} \frac{\sigma(\lambda-\mu)}{f(\lambda-\mu)} 
\{ B(\mu)E_{a}(\lambda) + F(\mu)C_{a}(\lambda) \},~ a,b,c = 1,2
\ear
where here and in the following  repeated indices denote the sum
operation. The elements of the vector 
${\vec \xi}$ are 
\EQ
{\vec \xi} = 
\matrix{(
0  &1  &-1  &0 )  \cr}
\EN
and the matrix $\tilde{r}(\lambda)$ has the structure
\EQ
\tilde{r}(\lambda) = \pmatrix{
1  &0  &0  &0  \cr
0  &s(\lambda)  &t(\lambda)  &0  \cr
0  &t(\lambda)  &s(\lambda)  &0  \cr
0  &0  &0  &1  \cr}.
\EN
Previous experience with the utilization of commutation relations 
for other systems \cite{FAD,QI,DV,DV1,RU} suggests us that the term
 $B(u)E_{a}(\lambda)$ has a wrong order in the commutation rule (18). This 
can be disentangled by 
commuting the operators $B(\mu)$ and $E_{a}(\lambda)$ with the help of the following additional 
commutation relations
\bear
B(\mu)E_{a}(\lambda) = 
\frac{f(\lambda-\mu)}{p(\lambda-\mu)} E_{a}(\lambda)B(\mu) + 
 \frac{m(\lambda-\mu)}{p(\lambda-\mu)}F(\lambda)C_{a}(\mu)
\nonumber \\
- \frac{\sigma(\lambda-\mu)}{p(\lambda-\mu)} \xi_{bc} 
 B_{b}(\mu)A_{ac}(\lambda) 
- \frac{n(\lambda-\mu)}{p(\lambda-\mu)} F(\mu)C_{a}(\lambda),~ a=1,2
\ear
obtaining as a final result the expression \footnote{Here we have used the identity $n(x) -p(x) = 1 $.}
\bear
A_{ab}(\lambda)B_{c}(\mu) = 
\frac{1}{f(\lambda-\mu)}{r}_{ed}^{bc}(\lambda-\mu)B_{e}(\mu)A_{ad}(\lambda) 
- \frac{m(\lambda-\mu)}{f(\lambda-\mu)} B_{b}(\lambda)A_{ac}(\mu) 
- \xi_{bc} \frac{\sigma(\lambda-\mu)}{f(\lambda-\mu)}  
\nonumber \\
\left \{ \frac{f(\lambda-\mu)}{p(\lambda-\mu)}E_{a}(\lambda)B(\mu) 
+\frac{m(\lambda-\mu)}{p(\lambda-\mu)}F(\lambda)C_{a}(\mu)  
- \frac{1}{p(\lambda-\mu)}F(\mu)C_{a}(\lambda) \right \}
,~ a,b,c = 1,2
\ear
where now the matrix $r$ has the following form
\EQ
r(\lambda) = 
\pmatrix{
1  &0  &0  &0  \cr
0  &b(\lambda)  &a(\lambda)  &0  \cr
0  &a(\lambda)  &b(\lambda)  &0  \cr
0  &0  &0  &1  \cr}
\EN
with
\EQ
a(\lambda) = \frac{\lambda}{\lambda+1},~~ b(\lambda) = \frac{1}{\lambda+1}
\EN
It is remarkable that such $r$-matrix is precisely 
the one  appearing on the isotropic $6$-vertex
(or the XXX Heisenberg) model. In some sense the Boltzmann 
weights of the $spl(2|1)$ $R$-matrix (2) conspire together in the
commutation rules in order to give us as a 
hidden symmetry the $6$-vertex structure. We think that
this is the algebraic reason why the 
bare scattering matrix of the equivalent fermionic problem 
(in the coordinate Bethe ansatz approach) \cite{FRA,KU} 
was previously determined to have the 
6-vertex form.
The other commutation rules concerning the operators $B_{a}(\lambda)$ which are important in the
diagonalization problem (6) can be obtained 
by basically following the method described above. They are
given by
\EQ
B(\lambda)B_{a}(\mu) = 
\frac{l(\mu-\lambda)}{f(\mu-\lambda)} B_{a}(\mu)B(\lambda) - 
\frac{m(\mu-\lambda)}{f(\mu-\lambda)} B_{a}(\lambda)B(\mu),~ a=1,2
\EN
\bear
D(\lambda)B_{a}(\mu) = 
\frac{g(\lambda-\mu)}{p(\lambda-\mu)} B_{a}(\mu)D(\lambda) 
+ \frac{q(\lambda-\mu)}{p(\lambda-\mu)} F(u)C_{a+3}(\lambda)
 \nonumber \\ 
 - \frac{n(\lambda-\mu)}{p(\lambda-\mu)} F(\lambda)C_{a+3}(\mu)
- \frac{\sigma(\lambda-\mu)}{p(\lambda-\mu)} 
\xi_{cb} E_{b}(\lambda)A_{ca}(\mu),~ a=1,2
\ear
\bear
B_{a}(\lambda)B_{b}(\mu) = \frac{1}{l(\lambda-\mu)}{r}_{cd}^{ab}
(\lambda-\mu)B_{c}(\mu)B_{d}(\lambda) 
-\xi_{ab} \frac{\sigma(\lambda-\mu)}{l(\lambda-\mu)p(\lambda-\mu)}  
\nonumber \\
\{ l(\lambda-\mu)F(\lambda)B(\mu) - F(\mu)B(\lambda) \},~ a=b=1,2 
\ear

The commutation rules (22,25-27) form the basis of 
the algebraic Bethe ansatz for the creation
operator $B_{a}(\lambda)$. It turns out, however, 
that $F(\lambda)$ also works as a creation operator,
and, as we shall see in next section, it plays an important role in the eigenstate construction. 
Therefore, the commutation relations with the operator $F(\lambda)$ are also necessary and they are
given by the expressions
\bear
A_{aa}(\lambda)F(\mu) = 
\frac{g(\lambda-\mu)}{f(\lambda-\mu)}
[1 - \frac{q^{2}(\lambda-\mu)}{g^{2}(\lambda-\mu)}] F(\mu)A_{aa}(\lambda)
+ \frac{q(\lambda-\mu)}{g(\lambda-\mu)}E_{a}(\lambda)B_{a}(\mu)
\nonumber \\
 - \frac{m(\lambda-\mu)}{f(\lambda-\mu)}B_{a}(\lambda)E_{a}(\mu) 
+ \frac{m(\lambda-\mu)}{g(\lambda-\mu)}\frac{q(\lambda-\mu)}
{f(\lambda-\mu)} F(\lambda)A_{aa}(\mu),~ a=1,2
\ear
\EQ
B(\lambda)F(\mu) = 
\frac{l(\mu-\lambda)}{p(\mu-\lambda)} F(\mu)B(\lambda) - 
\frac{n(\mu-\lambda)}{p(\mu-\lambda)} F(\lambda)B(\mu)
-  \frac{\sigma(\mu-\lambda)}{p(\mu-\lambda)} \xi_{ab} B_{a}(\lambda)B_{b}(\mu)
\EN
\EQ
D(\lambda)F(\mu) = 
\frac{r(\lambda-\mu)}{p(\lambda-\mu)} F(\mu)D(\lambda) - 
\frac{n(\lambda-\mu)}{p(\lambda-\mu)} F(\lambda)D(\mu)
+ \frac{\sigma(\lambda-\mu)}{p(\lambda-\mu)} \xi_{ab} E_{a}(\lambda)E_{b}(\mu)
\EN

Other necessary commutation relations are mentioned in the appendices $B$
and $C$. In the next section we shall
use all of them in the construction of the eigenstates $\ket{\Phi}$.

\section{The construction of the eigenstates and the eigenvalues}

We now have almost the complete machinery to start the construction of the 
eigenstates of the $spl(2|1)$ transfer matrix. The eigenstates can be obtained by acting
the creation operators $B_{a}(\lambda)$ and $F(\lambda)$ over the ferromagnetic 
pseudovacuum $\ket{0}$. Instead of presenting the general solution, we find more
illuminating first to discuss our method of construction for the lowest eigenstates. The first
excitation over the pseudovacuum $\ket{0}$, i.e. the one-particle state, is given in terms
of a linear combination between the operators $B_{a}(\lambda)$ by
\EQ
\ket{\Phi_{1}(\lambda_{1})} = B_{a}(\lambda_{1})F^{a}\ket{0}
\EN

The diagonalization problem (6) for the one-particle eigenstate (31) 
is solved by
making use of the commutation rules (22,25-27) and of properties 
(15,16), and we find the following
important relations
\EQ
B(\lambda) \ket{\Phi_{1}(\lambda_{1})} = 
\frac{l(\lambda_{1}-\lambda)}{f(\lambda_{1}-\lambda)} 
[l(\lambda)]^L \ket{\Phi_{1}(\lambda_{1})}
 - \frac{m(\lambda_{1}-\lambda)}{f(\lambda_{1}-\lambda)} [l(\lambda_{1})]^L B_{a}(\lambda)F^{a} \ket{0}
\EN
\EQ
D(\lambda) \ket{\Phi_{1}(\lambda_{1})} =
\frac{g(\lambda-\lambda_{1})}{p(\lambda-\lambda_{1})} [p(\lambda)]^L 
\ket{\Phi_{1}(\lambda_{1})}
+ \frac{\sigma(\lambda-\lambda_{1})}
{p(\lambda-\lambda_{1})} [f (\lambda_{1})]^L \xi_{ab}E_{a}(\lambda)F^{b} \ket{0}
\EN
\bear
\sum_{a=1}^{2}  A_{aa}(\lambda) \ket{\Phi_{1}(\lambda_{1})} = 
\frac{1}{f(\lambda-\lambda_{1})} [1+a(\lambda-\lambda_{1})]
[f(\lambda)]^L \ket{\Phi_{1}(\lambda_{1})}
\nonumber \\
  - \frac{m(\lambda-\lambda_{1})}{f(\lambda-\lambda_{1})} [f(\lambda_{1})]^L B_{a}(\lambda)F^{a} \ket{0} 
- \frac{\sigma(\lambda-\lambda_{1})}
{p(\lambda-\lambda_{1})} [l(\lambda_{1})]^L \xi_{ab} E_{a}(\lambda)F^{b} \ket{0} 
\ear

The terms proportional to the eigenstate $\ket{\Phi_{1}(\lambda_{1})}$ are 
denominated `wanted terms' and 
contribute for the eigenvalue $\Lambda(\lambda,\lambda_{1})$. The remaining ones are the so called 
`unwanted terms' and they must be canceled out. From 
expressions (32-34) we can see that this is the case, provided 
that \footnote{ Here it is interesting to point out that condition (35) 
does not impose any further restriction on the constants $F^a$ appearing
in the linear combination (31). As a consequence, the corresponding 
eigenvalue is double degenerated ( see also section (7) ).}
\EQ
\left [\frac{l(\lambda_{1})}{f(\lambda_{1})}\right ]^{L} = 1
\EN
where we have used the reflection property
\EQ
\frac{m(\lambda)}{f(\lambda)} = -\frac{m(-\lambda)}{f(-\lambda)}
\EN

The two-particle state $\ket{\Phi_{2}(\lambda_{1},\lambda_{2})}$ depends both of the operators $B_{a}(\lambda)$
and $F(\lambda)$. This becomes clear if we consider the commutation rule (27), suggesting the following ansatz
\EQ
\ket{\Phi_{2}(\lambda_{1},\lambda_{2})} = 
B_{a}(\lambda_{1})B_{b}(\lambda_{2})F^{ba}\ket{0} +
 h(\lambda_{1},\lambda_{2}) [l(\lambda_2)]^L F(\lambda_{1}) \xi_{ba}F^{ba} 
\ket{0}
\EN

The state $\ket{\Phi_{2}(\lambda_{1},\lambda_{2})}$ degenerates several unwanted terms if one tries to solve
the corresponding eigenvalue with the help of the commutation rules (22,25-27) 
and (28-30) for the operators $B_{a}(\lambda)$
and $F(\lambda)$, respectively. There exists some unwanted 
terms which can be automatically canceled out by an
appropriate choice of the function $h(\lambda_{1},\lambda_{2})$. For instance, the unwanted terms
\bear
[l(\lambda_{2})]^L \xi_{ab} E_{a}(\lambda)E_{b}(\lambda_{1}), ~~
[l(\lambda_{2})]^L B_{a}(\lambda)E_{a}(\lambda_{1})
\ear
are all of them proportional to $(F^{12} - F^{21})$ and can be excluded by fixing the following form for the 
function $h(\lambda_{1},\lambda_{2})$
\EQ
h(\lambda_{1},\lambda_{2}) = h(\lambda_1 -\lambda_2) =
 - \frac{\sigma(\lambda_{1}-\lambda_{2})}{p(\lambda_{1}-\lambda_{2})}
\EN

The other remaining unwanted terms need a further restriction in order to be canceled out. For example,
by collecting the contributions concerning the term of kind $B_{a}(\lambda)B_{b}(\lambda_{2})$ we find that
\EQ
-\frac{m(\lambda-\lambda_{1})}{f(\lambda-\lambda_{1})}
\left \{\frac{1}{f(\lambda_{1}-\lambda_{2})} [f(\lambda_{1})]^L
r_{ba}^{lm}(\lambda_{1}-\lambda_{2})F^{ml} - 
[l(\lambda_{1})]^L\frac{l(\lambda_{2}-\lambda_{1})}
{f(\lambda_{2}-\lambda_{1})}F^{ba} \right \}
B_{a}(\lambda)B_{b}(\lambda_{2}) \ket{0}
\EN

Considering the identity
\EQ
\frac{l(\lambda)}{f(\lambda)} = - \frac{1}{f(-\lambda)}
\EN
the term given in equation (40)  is canceled by imposing the following restriction
\EQ
\left [ \frac{l(\lambda_{i})}{f(\lambda_{i})} \right ]^{L}F^{ba} = 
- r_{ba}^{lm}(\lambda_i - \lambda_j)F^{ml},~~ i \neq j 
\EN

In fact, in Appendix $B$ we show that all unwanted terms of many different kinds can be canceled out by 
using such restrictions for both $\lambda_{1}$ and $\lambda_{2}$. We 
remark that conditions (42) are  particular 
cases of the general Bethe ansatz equations which 
are going to be discussed in the next section. Before going on, 
it is important to notice that a certain recurrence relation can 
be established between the one and the two-particle states. 
In order to see this, it is convenient to write
our results in a more compact way. Let us define the $n$-particle state as
\EQ
\ket{\Phi_{n}(\lambda_{1}, \cdots ,\lambda_{n})} =  
\vec {\Phi}_{n}(\lambda_{1}, \cdots ,\lambda_{n}).\vec{F} \ket{0}
\EN
where $\vec {\Phi}_{n}(\lambda_{1},\cdots,\lambda_{n})$ 
and $\vec{F}$ are vectors with $2^n$ components. Here we shall denote
the components of vector $\vec{F}$ by $F^{a_n,\cdots,a_1}$. Considering the
bilinear vector $\vec{B}(\lambda)$ as
\EQ
\vec {B}(\lambda) =
\pmatrix{ 
B_{1}(\lambda)  &B_{2}(\lambda) \cr}
\EN
and taking into account our previous results (31,37) and (39), the corresponding two-particle vector $\vec{\Phi}_{2}(\lambda_{1},\lambda_{2})$
is then written as
\EQ
\vec {\Phi}{_2}(\lambda_{1},\lambda_{2}) = 
\vec {B}(\lambda) \otimes \vec{\Phi}_{1}(\lambda_{2}) +
[l(\lambda_{2})]^L \frac{\sigma(\lambda_{1}-\lambda_{2})}{p(\lambda_{1}-\lambda_{2})} F(\lambda_{1}) \vec {\xi} \otimes \vec {\Phi}_{0}
\EN
where $\vec{\Phi}_{0}$ is the unitary constant.
A remarkable property present in equation (45) is the  symmetry 
under the exchange $\lambda_{1} \leftrightarrow 
\lambda_{2}$. More precisely one can show that
\EQ
\vec {\Phi}_{2}(\lambda_{1},\lambda_{2}) = 
\vec {\Phi}_{2}(\lambda_{2},\lambda_{1}) \frac{r_{12}(\lambda_{1}-\lambda_{2})}{l(\lambda_{1}-\lambda_{2})}
\EN
where we have used the commutation rules (27) and the identity 
\EQ
\frac{h(\lambda)}{h(-\lambda)} = r_{12}^{12}(\lambda) - r_{21}^{12}(\lambda)
\EN

The exchange symmetry between the variables ${\lambda_{i}}$ is always a welcome feature in the
algebraic Bethe ansatz analysis (see e.g. refs. \cite{DV,TA}). Such property can be used to cancel several kinds
of unwanted terms differing under the permutation on variables 
$\{\lambda_{i}\}$. By using this property we can avoid 
extra cumbersome mathematical analysis, like 
that appearing on the direct proof we gave in Appendix $B$ for the
variable $\lambda_{2}$ entering in the two-particle state. This last 
discussion and the recurrence relation (45) serve 
as a motivation for us  go on and to search for the three-particle 
eigenstate. As before, we start with an ansatz
which is able to collect together the `easy' unwanted terms. We find that the three-particle
state has the following structure
\bear
\vec{\Phi}(\lambda_{1},\lambda_{2},\lambda_{3}) =
\vec B(\lambda_{1}) \otimes \vec \Phi_{2}(\lambda_{2},\lambda_{3})
 +
[l(\lambda_{2})]^LF(\lambda_{1})\frac{\sigma(\lambda_{1}-\lambda_{2})}
{p(\lambda_{1}-\lambda_{2})} \vec{\xi} \otimes \vec \Phi_{1}(\lambda_{3}) 
\hat{F}_{2}(\lambda_{2},\lambda_{3})
\nonumber \\
+[l(\lambda_{3})]^LF(\lambda_{1})\frac{\sigma(\lambda_{1}-\lambda_{3})}
{p(\lambda_{1}-\lambda_{3})} \vec{\xi} \otimes 
\vec \Phi_{1}(\lambda_{2}) \hat{F}_{3}(\lambda_{2},\lambda_{3})
\ear

A simple way of determining the unknown function $\hat{F}_{j}(\lambda_{2},\lambda_{3})$ is by combining the exchange symmetry property  with
the direct cancellation of the simplest unwanted terms. From the permutation symmetry we find the
relations
\EQ
\hat{F}_{2}(\lambda_{2},\lambda_{3}) = \frac{l(\lambda_{3}-\lambda_{2})}{f(\lambda_{3}-\lambda_{2})} I
\EN
\EQ
\hat{F}_{3}(\lambda_{2},\lambda_{3}) = 
\frac{r_{23}(\lambda_{2}-\lambda_{3})}{f(\lambda_{2}-\lambda_{3})}
\EN

For completeness, in Appendix $C$ we have presented some details of this analysis as well
as how certain unwanted terms are easily canceled out. Now, after this detailed exposition of
our method, the general $n$-particle eigenstate can be, then, 
obtained as an induction of the
expressions (45) and (48). As a final result, we obtain that the $n$-particle state satisfies the following
recurrence relation
\bear
\vec {\Phi}_{n}(\lambda_{1},\cdots ,\lambda_{n}) = 
\vec {B}(\lambda_{1}) \otimes \vec {\Phi}_{n-1}(\lambda_{2}, 
\cdots ,\lambda_{n})
 + 
F(\lambda_{1})\vec{\xi} \otimes \sum_{j=2}^n [l
(\lambda_{j})]^L \frac{\sigma(\lambda_{1}-\lambda_{j})}
{p(\lambda_{1}-\lambda_{j})} 
\nonumber \\
\prod_{k=2,k \neq j}^{n} 
\frac{l(\lambda_{k}-\lambda_{j})}
{f(\lambda_{k}-\lambda_{j})} 
 \vec {\Phi}_{n-2}(\lambda_{2},\cdots,\lambda_{j-1},\lambda_{j+1},\cdots,\lambda_{n}) \prod_{k=2}^{j-1} \frac{r_{k,k+1}(\lambda_{k}-\lambda_{j})}{l(\lambda_{k}-\lambda_{j})}
\ear
where the index under the $r$-matrix denotes 
its action position on the tensor product $ \stackrel{1}{\otimes} \cdots 
\stackrel{k}{\otimes} \stackrel{k+1}{\otimes} 
\cdots \stackrel{n}{\otimes} $. Under
a consecutive permutation $\lambda_{j} \leftrightarrow \lambda_{j+1}$, the $n$-particle 
eigenstate satisfies the symmetry relation
\EQ
\vec {\Phi}_{n}(\lambda_{1},\cdots,\lambda_{j},\lambda_{j+1},\cdots,\lambda_{n}) = 
\vec {\Phi}_{n}(\lambda_{1},\cdots,\lambda_{j+1},\lambda_{j},\cdots,\lambda_{n}) 
\frac{r_{j,j+1}(\lambda_{j}-\lambda_{j+1})}{l(\lambda_{j}-\lambda_{j+1})}
\EN

Here we would like to make some remarks. It is interesting to notice that the 
recurrence relation (51) resembles in some way the one found by Tarasov \cite{TA} in the algebraic 
solution of the Izergin-Korepin model ($A_{2}^{2}$ system 
\cite{IK}). Of course, our case is 
much more involved due to the necessity of vectorial notation and  
the presence of a 
hidden symmetry enhanced by the vector 
$\vec \xi$ and the $6$-vertex $r$-matrix. Disregarding
the obvious subtleties of our construction, one is tempted 
to interpret (51) as the simplest
vectorial generalization of the one found by Tarasov for the $A_{2}^{2}$ model. In fact, the $6$-vertex 
model is one of the simplest system possessing a non-diagonal and factorizable $r$-matrix.
In this sense, we believe that our recurrence relation (51) is an important guideline for further
generalizations concerning the presence of other non-diagonal $r$-matrix. 

	In order to close this section, we turn to the restriction condition on the variables
$\{ \lambda_{i}\}$. An important unwanted term comes from the structure 
$B_{a_{1}}(\lambda_{1})B_{a_2}(\lambda_2) \cdots B_{a_{n}}(\lambda_{n})$ when one of the ${\lambda_{i}}$ is 
exchanged with $\lambda$, producing for example the 
$ B_{a_{1}}(\lambda)B_{a_2}(\lambda_2) \cdots B_{a_{n}}(\lambda_{n})$  unwanted term. Such term is generated by the
action of the operator $\sum_{a} A_{aa}(\lambda)$ and $B(\lambda)$. By using the commutation
rules (25) the term coming from $B(\lambda)$ is
\EQ
\frac{m(\lambda-\lambda_{1})}{f(\lambda-\lambda_{1})} [l(\lambda_{1})]^L
\prod_{j=2}^{n} \frac{l(\lambda_{j}-\lambda_{1})}
{f(\lambda_{j}-\lambda_{1})}F^{a_{n} \cdots a_{1}}
B_{a_{1}}(\lambda)B_{a_{2}}(\lambda_{2}) \cdots B_{a_{n}}(\lambda_{n}) \ket{0}
\EN
and under action of $\sum_{a} A_{aa}(\lambda)$ we get
\bear
-\frac{m(\lambda-\lambda_{1})}{f(\lambda-\lambda_{1})} [f(\lambda_{1})]^L
\prod_{j=2}^{n} \frac{1}{f(\lambda_{1}-\lambda_{j})}
r_{c_{1}d_{1}}^{a_{1}a_{2}}(\lambda_{1}-\lambda_{2})
r_{c_{2}d_{2}}^{d_{1}a_{3}}(\lambda_{1}-\lambda_{3}) \cdots 
r_{c_{n-1}d_{n-1}}^{d_{n-2}a_{n}}(\lambda_{1}-\lambda_{n})
\nonumber \\
F^{a_{n} \cdots a_{1}}
B_{d_{n-1}}(\lambda)B_{c_{1}}(\lambda_{2}) 
\cdots B_{c_{n-1}}(\lambda_{n}) \ket{0}
\ear

One can write this last term in a 
more compact form by defining an auxiliary transfer
matrix associated to the problem of an 
inhomogeneous $6$-vertex system as 
\EQ
T^{(1)}(\lambda,\{\lambda_{i}\})_{b_{1} \cdots b_{n}}^{a_{1} \cdots a_{n}} = 
r_{b_{1}d_{1}}^{c_{1}a_{1}}(\lambda-\lambda_{1})
r_{b_{2}c_{2}}^{d_{1}a_{2}}(\lambda-\lambda_{2}) \cdots
r_{b_{n}c_{1}}^{d_{n-1}a_{n}}(\lambda-\lambda_{n})
\EN
and equation (54)  becomes
\EQ
-\frac{m(\lambda-\lambda_{1})}{f(\lambda-\lambda_{1})} [f(\lambda_{1})]^L
\prod_{j=2}^{n} \frac{1}{f(\lambda_{1}-\lambda_{j})}
T^{(1)}(\lambda=\lambda_i,\{\lambda_{j} \})_{b_{1} \cdots b_{n}}^{a_{1} \cdots a_{n}}
F^{a_{n} \cdots a_{1}}
B_{b_{1}}(\lambda)B_{b_{2}}(\lambda_{2}) \cdots B_{b_{n}}(\lambda_{n}) \ket{0}
\EN

Here we remark that the same kind of reasoning can be done for any $\lambda_{i}$, 
namely if $B_{a_{i}}(\lambda)$ replaces 
$B_{a_{i}}(\lambda_{i})$. Now one has to use the 
property (52) and to 
perform cyclic permutations until 
one gets the variable $\lambda_{i}$ on the
first place. The basic difference is that now we get 
a string of ordered $r$-matrices multiplying the
components $F^{a_{n} \cdots a_{1}}$ of  vector $\vec{F}$. This 
argument is commonly used in many other algebraic constructions 
\cite{DV,TA} and it stresses the importance of the exchange symmetry (52). Taking
into account such discussion and collecting together 
equations (53) and (56), we find that the unwanted term 
$ B_{a_{1}}(\lambda_{1}) \cdots B_{a_{i}}(\lambda) 
\cdots B_{a_{n}}(\lambda_{n})$ is canceled provided 
that
\EQ
\left[ \frac{l(\lambda_{i})}{f(\lambda_{i})} \right]^{L}
\prod_{j \neq i}^{n} \frac{f(\lambda_{i}-\lambda_{j})
l(\lambda_{j}-\lambda_{i})}{f(\lambda_{j}-\lambda_{i})}
 F^{a_{n} \cdots a_{1}} =
 T^{(1)}(\lambda=\lambda_{i},\{\lambda_{j}\})_{a_{1} 
\cdots a_{n}}^{b_{1} \cdots b_{n}}
F^{b_{n} \cdots b_{1}}
\EN
which for $n=1,2$ reproduce our early conditions (35) and (42), respectively.

Other crucial unwanted 
term is that made by replacing one $B_{a_{i}}(\lambda)$ by
$E_{a_{i}}(\lambda)$, such as $E_{a_{1}}(\lambda)B_{a_{2}}(\lambda_{2})...B_{a_{n}}(\lambda_{n})$.
These terms come out the contribution of $\sum_{a} A_{aa}(\lambda)$ and $D(\lambda)$. The
piece which comes from $\sum_{a} A_{aa}(\lambda)$  has the form
\EQ
\frac{\sigma(\lambda-\lambda_{1})}{p(\lambda-\lambda_{1})}[l(\lambda_{1})]^L
\prod_{j=2}^{n} \frac{l(\lambda_{j}-\lambda_{1})}
{f(\lambda_{j}-\lambda_{1})}(-1)^{a_{1}}F^{a_{n} \cdots a_{1}}
E_{a_{1}}(\lambda)B_{a_{2}}(\lambda_{2}) \cdots B_{a_{n}}(\lambda_{n}) \ket{0}
\EN
and the contribution of $D(\lambda)$ is
\bear
-\frac{\sigma(\lambda-\lambda_{1})}{p(\lambda-\lambda_{1})} [f(\lambda_{1})]^L
\prod_{j=2}^{n} \frac{1}{f(\lambda_{1}-\lambda_{j})}
r_{c_{1}d_{1}}^{a_{1}a_{2}}(\lambda_{1}-\lambda_{2})
r_{c_{2}d_{2}}^{d_{1}a_{3}}(\lambda_{1}-\lambda_{3}) \cdots
r_{c_{n-1}d_{n-1}}^{d_{n-2}a_{n}}(\lambda_{1}-\lambda_{n})
\nonumber \\
F^{a_{n} \cdots a_{1}}
\xi_{lk} E_{k}(\lambda)B_{c_{1}}(\lambda_{2}) \cdots B_{c_{n-1}}(\lambda_{n}) \delta_{l,d_{n-1}} \ket{0}
\ear

By combining these two terms together and by using 
definition (55) one concludes that such unwanted
terms are canceled out by the same restriction (57), as it should be. Of course we have several others
non-trivial
\footnote{Some trivial unwanted terms come 
in pairs and are automatically eliminated
such as we have exemplified for the 
cases of the two and three-particle states.}
 unwanted terms.
Although we do not have a systematic proof 
that such terms vanish, the checks we performed 
so 
far have been rather exhaustive ( see e.g. appendices $B$ and $C$ ), 
and seem to us to establish beyond reasonable doubt that the
restriction (57) is the unic condition to be imposed 
on the $n$-particle eigenstate in order to cancel the
unwanted terms.

Finally, the eigenvalue $\Lambda(\lambda,\{\lambda_{i}\})$ of the $n$-particle eigenstate
can be calculated by keeping only the terms proportional to
 the eigenstate $\ket{ {\Phi}_{n}(\lambda_{1},\cdots,\lambda_{n})}$. For 
instance, that
proportional to the vector 
$B_{a_{1}}(\lambda_{1})B_{a_{2}}(\lambda_{2}) \cdots B_{a_{n}}(\lambda_{n})$
is determined by the extensive use of the first terms of the
commutation rules (22,25-27), and we find that
\EQ
\Lambda(\lambda,\{\lambda_{i}\}) = 
[f(\lambda)]^L \prod_{i=1}^{n} \frac{1}{f(\lambda-\lambda_{i})} 
\Lambda^{(1)}(\lambda,\{\lambda_{i}\}) + 
[l(\lambda)]^L \prod_{i=1}^{n} \frac{l(\lambda_{i}-\lambda)}{f(\lambda_{i}-\lambda)} + 
[p(\lambda)]^L \prod_{i=1}^{n} \frac{g(\lambda-\lambda_{i})}{f(\lambda-\lambda_{i})}  
\EN
where $\Lambda^{(1)}(\lambda,\{\lambda_{i}\})$ is the eigenvalues of the auxiliary problem related to
the transfer matrix of the inhomogeneous $6$-vertex model, i.e.
\EQ
 T^{(1)}(\lambda,\{\lambda_{i}\})_{a_{1} \cdots a_{n}}^{b_{1} \cdots b_{n}}
F^{b_{n} \cdots b_{1}} = \Lambda^{(1)}(\lambda,\{\lambda_{i}\})  
F^{a_{n} \cdots a_{1}}
\EN

In conclusion, the results of this section show us that the computation of the 
eigenstates and the eigenvalues of the $spl(2|1)$ model is  still dependent 
of an additional
 eigenvalue problem for the inhomogeneous transfer matrix 
$ T^{(1)}(\lambda,\{\lambda_{i}\})$. We shall discuss 
this matter in the next section.

\section{The nested Bethe ansatz equations}

The auxiliary problem (61) can still be solved 
by an algebraic approach since  the
Yang-Baxter algebra (7) is also valid for 
inhomogeneous transfer matrices. The corresponding
monodromy matrix has the form
\EQ
{\cal T}^{(1)}(\lambda,\{\lambda_{i}\}) = 
{\cal L}^{(1)}_{on}(\lambda-\lambda_{n}){\cal L}^{(1)}_{on-1}
(\lambda-\lambda_{n-1}) \cdots {\cal L}^{(1)}_{o1}(\lambda-\lambda_{1})
\EN
where ${\cal L}^{(1)}(\lambda)$ is the vertex operator 
of the isotropic $6$-vertex model and its
matrix elements on the space $C^{2} \times C^{2}$ are given by
\EQ
{\cal L}^{(1)}(\lambda) = 
\pmatrix{
1  &0  &0  &0  \cr
0  &a(\lambda)  &b(\lambda)  &0  \cr
0  &b(\lambda)  &a(\lambda)  &0  \cr
0  &0  &0  &1  \cr}
\EN

In order to go on, we basically have to adapt 
the well known algebraic results for the
homogeneous $6$-vertex model of Faddeev et al \cite{FAD,QI,DV} in the case of an irregular lattice. We
recall that this problem has appeared 
in many different contexts in the literature \cite{DV1,RU}, but for sake of
completeness we present the main results. Taking 
the monodromy matrix as
\EQ
{\cal T}^{(1)}(\lambda,\{\lambda_{i}\}) =
\pmatrix{
A^{(1)}(\lambda,\{\lambda_{i}\})  &  B^{(1)}(\lambda,\{\lambda_{i}\})  \cr
C^{(1)}(\lambda,\{\lambda_{i}\})  &  D^{(1)}(\lambda,\{\lambda_{i}\})  \cr}
\EN
and defining the reference state as
\EQ
\ket{0^{(1)}} = \prod_{i=1}^{n} \otimes \ket{0^{(1)}}_i,~~
\ket{0^{(1)}}_i =
\pmatrix{
1  \cr
0  \cr}
\EN
we find the following properties
\EQ
A^{(1)}(\lambda,\{\lambda_{i}\})\ket{0^{(1)}} = \ket{0^{(1)}},~~
D^{(1)}(\lambda,\{\lambda_{i}\})\ket{0^{(1)}} = \prod_{i=1}^{n} 
a(\lambda-\lambda_{i})\ket{0^{(1)}},~~
C^{(1)}(\lambda,\{\lambda_{i}\})\ket{0^{(1)}} = 0
\EN

The algebraic Bethe ansatz developed by Faddeev et al 
\cite{FAD,QI,DV} says that all eigenvectors of the
transfer matrix $A^{(1)}(\lambda,\{\lambda_{i}\}) + 
D^{(1)}(\lambda,\{\lambda_{i}\})$ can be written
as
\EQ
\ket{\Phi^{(1)}(\mu_{1},\cdots,\mu_{m})} = 
\prod_{i=1}^{m} B^{(1)}(\mu_{i},\{ \lambda_j \} ) \ket{0^{(1)}}
\EN
for some additional restriction on the variables $\{\mu_{i}\}$. Moreover, 
the Yang-Baxter algebra (7) 
for the monodromy matrix (62) and the $6$-vertex 
$r$-matrix (23) yields the following commutation
relations
\EQ
A^{(1)}(\lambda,\{\lambda_{i}\})B^{(1)}(\mu,\{\lambda_{i}\}) =
\frac{1}{a(\mu-\lambda)}B^{(1)}(\mu,\{\lambda_{i}\})
A^{(1)}(\lambda,\{\lambda_{i}\})
- \frac{b(\mu-\lambda)}{a(\mu-\lambda)}B^{(1)}(\lambda,\{\lambda_{i}\})
A^{(1)}(\mu,\{\lambda_{i}\}) 
\EN
\EQ
D^{(1)}(\lambda,\{\lambda_{i}\})B^{(1)}(\mu,\{\lambda_{i}\}) =
\frac{1}{a(\lambda-\mu)}B^{(1)}(\mu,\{\lambda_{i}\})
D^{(1)}(\lambda,\{\lambda_{i}\})
- \frac{b(\lambda-\mu)}{a(\lambda-\mu)}B^{(1)}(\lambda,\{\lambda_{i}\})
D^{(1)}(\mu,\{\lambda_{i}\}) 
\EN
\EQ
\left [ B^{(1)}(\mu,\{\lambda_{i}\}),B^{(1)}(\lambda,\{\lambda_{i}\})
\right ] = 0
\EN

By using the commutation relations (68-70) 
we can carry $A^{(1)}(\lambda,\{\lambda_{i}\}) + 
D^{(1)}(\lambda,\{\lambda_{i}\})$ 
through all $B^{(1)}(\mu_{i},\{ \lambda_j \})$, and we find that the eigenvalue is
\EQ
\Lambda^{(1)}(\lambda,\{\lambda_{i}\},\{\mu_{j}\}) =
\prod_{j=1}^{m} \frac{1}{a(\mu_{j}-\lambda)} +
\prod_{i=1}^{n} a(\lambda - \lambda_{i}) \prod_{j=1}^{m} \frac{1}{a(\lambda-\mu_{j})}
\EN
if the numbers $\{\mu_{j}\}$ satisfy the following system of equations
\EQ
\prod_{i=1}^{n} a(\mu_{j} - \lambda_{i}) = 
- \prod_{k=1}^{m} \frac{a(\mu_{j}-\mu_{k})}{a(\mu_{k}-\mu_{j})},~~
j=1,\cdots,m
\EN

Hence, by using the expression for $\Lambda^{(1)}(\lambda,\{\lambda_{i}\},
\{\mu_{j}\})$ in 
equation (60), we finally find that the eigenvalues of the $spl(2|1)$ model has the following 
general form
\bear
\Lambda(\lambda,\{\lambda_{i}\},\{\mu_{j}\}) = 
[f(\lambda)]^L \prod_{i=1}^{n} \frac{1}{f(\lambda-\lambda_{i})}
 \left \{ \prod_{j=1}^{m} \frac{1}{a(\mu_{j}-\lambda)} +
\prod_{i=1}^{n} a(\lambda - \lambda_{i}) 
\right . \nonumber \\ \left .
\prod_{j=1}^{m} \frac{1}{a(\lambda-\mu_{j})} \right \} +
[l(\lambda)]^L \prod_{i=1}^{n} 
\frac{l(\lambda_{i}-\lambda)}{f(\lambda_{i}-\lambda)} + 
[p(\lambda)]^L 
\prod_{i=1}^{n} \frac{g(\lambda-\lambda_{i})}{f(\lambda-\lambda_{i})}  
\nonumber \\
\ear
and the restriction condition (57) for the eigenstates becomes
\EQ
\left [ \frac{l(\lambda_{i})}{f(\lambda_{i})} \right ]^{L} =
-(-1)^{n} \prod_{l=1}^{m} \frac{1}{a(\mu_{l}-\lambda_{i})} 
\EN
where we have used the identity given in equation (41).

The set of coupled equations (72) and (74) are usually denominated nested Bethe
ansatz equations. They can be written in a more symmetric way if we perform the 
transformation
\EQ
\lambda_{i} \rightarrow \frac{\lambda_{i}}{2} - \frac{b-1/2}{2}, ~~
\mu_{j} \rightarrow \frac{\mu_{j}}{2} - \frac{b+1/2}{2}
\EN
and afterwords by performing the rescaling 
\EQ
\lambda_{i} \rightarrow \frac{\lambda_{i}}{i},~~
  \mu_{j} \rightarrow \frac{\mu_{j}}{i}
\EN
we get 
\begin{displaymath}
{\left [ \frac{\lambda_{i}-(b-1/2)i}{\lambda_{i}+(b-1/2)i}\right ] }^{L} = 
- (-1)^{L-n}
 \prod_{l=1}^{m} \frac{\lambda_{i}-\mu_{l} + i}{\lambda_{i} -\mu_{l} - i},
~i=1, \cdots, n
\end{displaymath}
\EQ
\prod_{l=1}^{m} \frac{\mu_{k} - \mu_{l} - 2i}{\mu_{k} - \mu_{l} + 2i} =
- \prod_{j=1}^{n} \frac{\mu_{k} - \lambda_{j} - i}{\mu_{k} - \lambda_{j} + i},~~k=1, \cdots, m 
\EN

The eigenvalues $\Lambda(\lambda,\{\lambda_{i}\},\{\mu_{l}\})$, 
after performing
the transformations (75) and (76), can be written as
\bear
\Lambda(\lambda,\{\lambda_{i}\},\{\mu_{l}\}) = 
\left [\frac{\lambda}{(1/2-b)i+\lambda}\right ]^{L} \left \{
 \prod_{i=1}^{n}\frac{\lambda_{i} + 
(b-1/2)i - 2\lambda}{\lambda_{i} - (b-1/2)i 
- 2\lambda}\prod_{l=1}^{m} \frac{\mu_{l} + 
(3/2-b)i - 2\lambda}{\mu_{l} - (b+1/2)i - 2\lambda} +
\right . \nonumber \\ \left .
\prod_{i=1}^{n}\frac{\lambda_{i} + (b-1/2)i - 
2\lambda}{\lambda_{i} - (b+3/2)i - 2\lambda} 
\prod_{l=1}^{m} \frac{\mu_{l} - (b+5/2)i - 
2\lambda}{\mu_{l} - (b+1/2)i - 2\lambda}
\right \} 
+\left [\frac{(1/2-b)i - \lambda}{(1/2-b)i + \lambda}\right ]^{L}
 \prod_{i=1}^{n} -\frac{\lambda_{i} + (b-1/2)i 
- 2\lambda}{\lambda_{i} - (b-1/2)i - 2\lambda} +
\nonumber \\
\left [\frac{\lambda(i+\lambda)}{[(1/2+b)i+\lambda]
[(1/2-b)i+\lambda]}\right ]^{L}
\prod_{i=1}^{n} -\frac{\lambda_{i} + (b-1/2)i - 2\lambda}{\lambda_{i} - (b+3/2)i - 2\lambda} \nonumber \\
\ear

In particular, the eigenenergies $E(L)$ of the corresponding Hamiltonian (10)
can be calculated by taking the logarithmic derivative of 
$\Lambda(\lambda,\{\lambda_{i}\},\{\mu_{l}\})$ at the regular point $\lambda=0$. More
precisely, by performing the operation $\left. i\frac{dln\Lambda(\lambda,\{
\lambda_i \},\{ \mu_j \})}{d\lambda} \right|_{\lambda=0}$
we find
\EQ
E(L) =\sum_{i=1}^{n} \frac{4(b-1/2)}{(\lambda_j^1)^2 +(b-1/2)^{2}} + 
\frac{2L}{b-1/2}
\EN
  
In order to conclude this section we would like to make the following remarks.
First of all, our analytical result (78) can be confronted 
with the conjecture made by 
Maassarani \cite{MA} concerning the 
structure of the eigenvalues of the $spl(2|1)$ 
model.  Here, one first has to disregard the presence of certain phase factors,
which is the basic difference between the standard and the supersymmetric 
formulation. We conclude that our analytical result for the eigenvalue 
$\Lambda(\lambda, \{ \lambda_i \}, \{ \mu_j \} ) $ is in agreement with the
conjecture made by Maassarani \cite{MA}, if one takes the rational limit of
equation (73) of ref. \cite{MA}. 
Our second remark is concerned with the extension of the
Bethe ansatz results obtained in this 
section for a more general classes of twisted boundary 
conditions. These boundary conditions 
correspond to the introduction of a seam
with different Boltzmann weights along the infinite direction on the cylinder. 
Such weights depend on two angles $\theta_{1}$ and $\theta_{2}$\footnote{In the
fermionic version \cite{FRA,KU} of the corresponding 
Hamiltonian (10) this means that the
fermions with spins up $c_{\uparrow}(i)$ and down $c_{\downarrow}(i)$ 
satisfy the boundary conditions $c_{\uparrow,\downarrow}(1) = 
e^{i\theta_{1,2}}c_{\uparrow,\downarrow}(L+1)$,
$c^{\dag}_{\uparrow,\downarrow}(1)) = 
e^{-i\theta_{1,2}}c^{\dag}_{\uparrow,\downarrow}(L+1)$} and we 
represent them by
the operator $\tilde{{\cal  L}}(\lambda)$ given by
\EQ
{\scriptsize
\tilde{{\cal L}}(\lambda) = 
\pmatrix{
 l(\lambda) &0 &0 &0 &0  &0  &0  &0  &0  &0  &0  &0  &0  &0  &0  &0    \cr
 0 & \omega_1 f(\lambda) &0 &0 &m(\lambda)  &0  &0  &0  &0  &0  &0  &0  &0  &0  &0  &0    \cr
 0 &0 &\omega_2 f(\lambda) &0 &0  &0  &0  &0  &m(\lambda)  &0  &0  &0  &0  &0  &0  &0    \cr
 0 &0 &0 &\omega_1 \omega_2 p(\lambda) &0  &0  &-\omega_2 \sigma(\lambda)  &0  &0  &\omega_1 \sigma(\lambda)  &0  &0  &n(\lambda)  &0  &0  &0    \cr
 0 &m(\lambda) &0 &0 &\frac{f(\lambda)}{\omega_1}  &0  
&0  &0  &0  &0  &0  &0  &0  &0  &0  &0    \cr
 0 &0 &0 &0 &0  &1  &0  &0  &0  &0  &0  &0  &0  &0  &0  &0    \cr
 0 &0 &0 &-\omega_2 \sigma(\lambda) &0  &0  &\omega_1\omega_2 t(\lambda)  &0  &0  &s(\lambda)  &0  &0  &-\frac{\sigma(\lambda)}{\omega_1}  &0  &0  &0    \cr
 0 &0 &0 &0 &0  &0  &0  &\omega_2 g(\lambda)  &0  &0  &0  &0  &0  &q(\lambda)  &0  &0    \cr
 0 &0 &m(\lambda) &0 &0  &0  &0  &0  &\frac{f(\lambda)}{\omega_2}  &0  &0  &0  &0  &0  &0  &0    \cr
 0 &0 &0 &\omega_1 \sigma(\lambda) &0  &0  &s(\lambda)  &0  &0  &\frac{t(\lambda)}{\omega_1\omega_2}  &0  &0  &\frac{\sigma(\lambda)}{\omega_2}  &0  &0  &0    \cr
 0 &0 &0 &0 &0  &0  &0  &0  &0  &0  &1  &0  &0  &0  &0  &0    \cr
 0 &0 &0 &0 &0  &0  &0  &0  &0  &0  &0  &\omega_1 
g(\lambda)  &0  &0  &q(\lambda)  &0    \cr
 0 &0 &0 &n(\lambda) &0  &0  &-\frac{\sigma(\lambda)}{\omega_1}  
&0  &0  &\frac{\sigma(\lambda)}{\omega_2}  &0  &0  
&\frac{p(\lambda)}{\omega_1 \omega_2}  &0  &0  &0    \cr
 0 &0 &0 &0 &0  &0  &0  &q(\lambda)  &0  &0  &0  &0  &0  &\frac{g(\lambda)}{\omega_2}  &0  &0    \cr
 0 &0 &0 &0 &0  &0  &0  &0  &0  &0  &0  &q(\lambda)  &0  &0  &\frac{g(\lambda)}{\omega_1}  &0    \cr
 0 &0 &0 &0 &0  &0  &0  &0  &0  &0  &0  &0  &0  &0  &0  &r(\lambda)    \cr}
}
\EN
where $ \omega_{1,2} = e^{-i \theta_{1,2}} $. 
The integrability is still preserved and the nested Bethe ansatz
equations (77) are modified in the following way
\begin{displaymath}
{\left [\frac{\lambda_{i}-(b-1/2)i}{\lambda_{i}+(b-1/2)i}\right ]}^{L} = 
- (-1)^{L-n} e^{-i\theta_{1}}
 \prod_{l=1}^{m} \frac{\lambda_{i}-\mu_{l} + i}{\lambda_{i} -\mu_{l} - i},
~i=1,\cdots,n
\end{displaymath}
\EQ
\prod_{l=1}^{m} \frac{\mu_{k} - \mu_{l} - 2i}{\mu_{k} - \mu_{l} + 2i} =
- e^{-i(\theta_{1}-\theta_{2})}\prod_{j=1}^{n} \frac{\mu_{k} - \lambda_{j} - i}{\mu_{k} - \lambda_{j} + i}, ~~
k= 1,\cdots,m
\EN
and the eigenenergy equation (78) remains unchanged.
One interesting case of twisted boundary conditions is when $\theta{_1} = 
-\theta_{2} = \vartheta/2$. In this case the $spl(2|1)$ algebra 
is preserved and is 
accomplished in terms of the $spl(2|1)$ generators as follows. The odd
generators transform as
\EQ
V_{\pm} \rightarrow e^{\pm \vartheta/2} V_{\pm} , ~~
 \overline {V}_{\pm} \rightarrow e^{\pm \vartheta/2} \overline{V}_{\pm}
\EN
and the even generators  behave as
\EQ
S_{\pm} \rightarrow e^{\pm \vartheta} S_{\pm}, ~~
S_{3} \rightarrow S_{3}, ~~
B \rightarrow B
\EN

This special boundary condition is quite useful, because by varying the
angle $\vartheta$  until 
$\vartheta = 2\pi$\footnote{Notice that this corresponds to
a periodic (anti-periodic) boundary condition for the bosonic (fermionic) 
degrees of freedom.} (in the sector where $(L-n)$ is even) we can study the
spectrum of the system in the supersymmetric formulation. In this sense, both the
standard and the supersymmetric formulation may be related by
performing in  determined sectors of theory a twisted boundary condition. 
This idea has been used with success to 
study the spectrum of the $OSP(1|2)$ spin chain \cite{OSP} .

The final remark consists of a discussion concerning the reflection
symmetry $b \rightarrow -b$ in terms of our algebraic formulation. As has 
already been noticed by Maassarani \cite{MA}, the $R$-matrix 
structure (2) allowed an 
additional reference state such as
\EQ
\ket{\tilde {0}} = \prod_{i=1}^{L} \otimes \ket{\tilde {0}}_{i} , ~~
\ket{\tilde {0}}_{i} =
\pmatrix{
0 \cr
0 \cr
0 \cr
1 \cr}
\EN

The triangular form of the vertex operator ${\cal L}(\lambda)$ is now
given by
\EQ 
{\cal L}(\lambda)\ket{\tilde{0}}_{i} =
\pmatrix{
p(\lambda)  &  0  &  0  &  0  \cr
*  &  g(\lambda)  &  0  &  0  \cr
*  &  0  &  g(\lambda)  &  0  \cr
*  &  *  &  *  &  r(\lambda)  \cr}
\ket{\tilde{0}}_i
\EN
which can be  related to the earlier triangular property (12) by $b \rightarrow -b$ and  $f_{1} \leftrightarrow f_{2}$.
Such relation becomes even more rigorous if one looks for the commutation rules
comming from the Yang-Baxter algebra (7). In this case, we find that it is 
more convenient to start with the 
following matrix form of the monodromy
$\tilde{{\cal T}}(\lambda)$
\EQ
\tilde{{\cal T}}(\lambda) =
\pmatrix{
\tilde{D}(\lambda)   &   \tilde{C}_{5}(\lambda)   &  \tilde{C}_{4}(\lambda)   &   \tilde{C}_{3}(\lambda)   \cr
\tilde{E}_{2}(\lambda)  & \tilde {A}_{22}(\lambda)  & \tilde{A}_{21}(\lambda)  &  \tilde{C}_{2}(\lambda)   \cr
\tilde{E}_{1}(\lambda)  &  \tilde{A}_{12}(\lambda)  &  \tilde{A}_{11}(\lambda)  &  \tilde{C}_{1}(\lambda)   \cr
\tilde{F}(\lambda)  &  \tilde{B}_{2}(\lambda)  &  \tilde{B}_{1}(\lambda)  &  \tilde{B}(\lambda)  \cr}
\EN

By using the Yang-Baxter algebra with this structure, the commutation
rules of the diagonal terms with operator  $B_a(\lambda)$, for example, 
are given by
\bear
\tilde {A}_{ab}(\lambda)\tilde {B}_{c}(\mu) = 
\frac{1}{g(\lambda-\mu)}{r}_{ed}^{bc}(\lambda-\mu)\tilde {B}_{e}(\mu) 
\tilde {A}_{ad}(\lambda) 
-\frac{q(\lambda-\mu)}{g(\lambda-\mu)} 
\tilde {B}_{b}(\lambda)\tilde {A}_{ac}(\mu)  +
\nonumber \\ 
\frac{\sigma(\lambda-\mu)}{g(\lambda-\mu)} \xi_{bc} 
\left \{ \frac{g(\lambda-\mu)}{p(\lambda-\mu)}E_{a}(\lambda)B(\mu) 
+\frac{q(\lambda-\mu)}{p(\lambda-\mu)}
\tilde {F}(\lambda)\tilde {C}_{a}(u)  
 - \frac{1}{p(\lambda-\mu)} \tilde {F}(\mu)
\tilde {C}_{a}(\lambda) \right \}
,~~a,b,c = 1,2 \nonumber \\
\ear
\EQ
\tilde {B}(\lambda)\tilde {B}_{a}(\mu) = 
\frac{r(\mu-\lambda)}{g(\mu-\lambda)} \tilde {B}_{a}(\mu)
\tilde {B}(\lambda) - \frac{q(\mu-\lambda)}{g(\mu-\lambda)} 
\tilde {B}_{a}(\lambda)\tilde {B}(\mu)
\EN
\bear
\tilde {D}(\lambda)\tilde {B}_{a}(\mu) = 
\frac{f(\lambda-\mu)}{p(\lambda-\mu)} \tilde {B}_{a}(\mu)
\tilde {D}(\lambda) + \frac{m(\lambda-\mu)}{p(\lambda-\mu)} 
\tilde {F}(\mu)\tilde{C}(\lambda)_{a+3}
\nonumber \\
 - \frac{n(\lambda-\mu)}{p(\lambda-\mu)} \tilde {F}(\lambda)\tilde{C}(\mu)_{a+3}
+ \frac{\sigma(\lambda-\mu)}{p(\lambda-\mu)} \xi_{cb}\tilde 
{E}_{b}(\lambda)\tilde {A}_{ca}(\mu)
\ear

Such commutation relations are the same of that given in equations (22,25,26)
by performing the reflection symmetry $b \rightarrow -b$ and $\sigma \rightarrow
-\sigma$. We have checked that similar situation appears for the other 
commutations rules. Since $\sigma \rightarrow -\sigma$ can be considered as
 a canonical 
transformation, we conclude that the eigenvalues 
parametrized by equations (77)
and (78) either with $+b$ or with $-b$, should produce the same spectrum for the
$spl(2|1)$ model. From the Bethe ansatz point of view, one would think that this
sounds a bit strange, since the topology of the nested Bethe ansatz equations 
are very different for $\pm b$. The correct interpretation is as follows. They
in fact can produce the same eigenvalues, but of course with different 
structure of zeros. Actually, we have checked 
this fact by solving numerically the nested
Bethe ansatz equation for some values of $\pm b$ with a finite lattice size L.

\section{The ground state structure}

The purpose of this section is to study both numerically and analytically
 the behaviour of the ground state of 
the $spl(2|1)$ Hamiltonian (10). Due to the
reflection symmetry $b \rightarrow -b$, it is enough to investigate the 
regimes $0 \leq b < 1/2$ and $1/2 < b < \infty$. We begin our analysis by
investigating the behaviour around the reference states with ferromagnetic
feature. In order to keep our previous notation (section 2) we define the 
four basic states acting on the $i$-th site of the one-dimensional lattice by
\EQ
f_{1}(i) = \pmatrix{
1  \cr
0  \cr
0  \cr
0  \cr}_{i},~~
b_{1}(i) = \pmatrix{
0  \cr
1  \cr
0  \cr
0  \cr}_{i},~~
b_{2}(i) = \pmatrix{
0  \cr
0  \cr
1  \cr
0  \cr}_{i},~~
f_{2}(i) = \pmatrix{
0  \cr
0  \cr
0  \cr
1  \cr}_{i}
\EN

The low-lying excitations over the ferromagnetic state
\EQ
\ket{\phi^{f_{1}}_{0}} = \ket{f_{1}(1) \cdots f_{1}(L)}
\EN
can be constructed if one takes the linear combination of the states made by
replacing one state $f_{1}(k)$ by $b_{1,2}(k)$, namely
\EQ
\ket{\phi^{f_{1}}_{1,2}} = 
\sum_{k=1}^{L} \alpha(k) \ket{f_{1}(1) \cdots b_{1,2}(k) \cdots f_{1}(L)}
\EN

While the eigenvalue of the ferromagnetic state is trivially determined to be
\EQ
E^{f_{1}}_{0}(b) = - \frac{2L}{1/2-b}
\EN
that of the low-lying excitation $\ket{\phi^{f_{1}}_{1,2}}$ is related to the solution
of the equation
\EQ
E^{f_{1}}_{1,2} \alpha(k) = -\frac{2}{1/2-b}(L-1)\alpha(k) + 
\frac{1}{1/2-b}[\alpha(k+1)+\alpha(k-1)]
\EN

This relation can be solved by taking $\alpha(k) = Ae^{ikp}$ and the
dispersion relation is
\EQ
E^{f_{1}}_{1,2}(p) = 
-\frac{2L}{1/2-b} + \frac{4}{1/2-b}\cos^{2}(p/2)
\EN
where under periodic boundary condition the momenta p assumes the values
\EQ
p = \frac{2\pi n}{L} ; ~~ n= 1,\cdots,L-1
\EN

Analogously, by repeating the same reasoning to the other ferromagnetic
state

\EQ
\ket{\phi^{f_{2}}_{0}} = \ket{f_{2}(1) \cdots f_{2}(L)}
\EN
we find the following results
\EQ
E^{f_{2}}_{0}(b) = - \frac{2L}{1/2+b}
\EN
\EQ
E^{f_{2}}_{1,2}(p) = 
-\frac{2L}{1/2+b} + \frac{4}{1/2+b}\cos^{2}(p/2)
\EN

In the regime $1/2 < b < \infty$ we have $E^{f_{2}}_{0}(b) < 0$ and 
$E^{f_{1}}_{0}(b) > 0$. Moreover, the sign in the dispersion relation of 
$E^{f_{2}}_{1,2}(p)$($E^{f_{1}}_{1,2}(p))$ is positive (negative). This 
suggests that the ground state is given by $E^{f_{2}}_{0}(b)$ and the excitations
over it grows until reaching the upper-bound $E^{f_{1}}_{0}(b)$. Remarkably 
enough, the statement concerning the ground state can be put in a more rigorous
ground. In fact, following Bader and Schilling \cite{BS} 
the ground state satisfies the
relation
\EQ
E_{0} \geq LE_{0}^{2}
\EN
where $E_{0}^{2}$ is the lowest eigenvalue of the two-body Hamiltonian 
 $H_{i,i+1}$. In the regime $0 < b < 1/2$ we find that
\EQ
E_{0}^{2} = - \frac{2}{1/2+b}
\EN

On the other hand, the expectation value of the $spl(2|1)$ Hamiltonian 
in the ferromagnetic state $\ket{f_{2}(1) \cdots f_{2}(L)}$ is $ - \frac{2L}{1/2+b}$,
and by the variational principle we have
\EQ
E_{0} \leq - \frac{2L}{1/2+b}
\EN

By comparing equations (100) and (102) we conclude that in the regime 
$1/2 < b < \infty$ the ground state energy is $E^{f_{2}}_{0}(b)$. Actually, 
by analysing
the dispersions relation $E^{f_{2}}_{1,2}(p)$ around $p \sim \pi$ we find that it is 
three-fold degenerated. The rest of the statement, i.e. that 
the spectrum satisfies
$-\frac{2}{1/2+b} \leq E/L \leq -\frac{2}{1/2-b}$, has been confirmed by a numerical diagonalization
of the Hamiltonian (10) up to $L \leq 12$.

In the regime $0 \leq b < 1/2$ rigorous results become more involved since
the Hamiltonian (10) is not hermitian \footnote{ Nevertheless, we have
numerically checked  up to $L \leq 12 $ that the ground state
is real. }. For instance, such `nice' argument of Bader and
Schilling \cite{BS} needs further elaboration. In this case, the ferromagnetic state
 $\ket{\phi^{f_{1}}_{0}}$ has lower energy than the state $\ket{\phi^{f_{2}}_{0}}$ and the sign on
the dispersion relation of $E^{f_{1}}_{1,2}(p)$ is now positive. This 
may suggest
that there exist states with much lower energy. Indeed, for $b=0$ we have found 
previously that the ground state has an antiferromagnetic structure and in the 
thermodynamic limit the energy per site $e_{\infty}$ was determined to be $e_{\infty}=
-8ln(2) \simeq -5.5452 < E_0^{f_1,f_2}(b=0)=-4. $.
Furthermore, the numerical diagonalization of Hamiltonian (10) for $b \neq 0$
$(0 \leq b < 1/2)$ shows that the ground state jumps into the many possible $U(1)$
sectors of the theory if one varies the lattice size L. 
Let us illustrate this from 
the Bethe ansatz point of view. As an example, let us assume that for sufficient small b some states have the zeros structure similar to that previously found for the ground
state at $b=0$ \cite{MP}, namely
\EQ
\lambda_{j} = \xi_{j} + i + O(e^{-aL}) , ~~ \mu_{j} = \xi
\EN

By substituting such structure of zeros in the nested Bethe ansatz equations,
(see ref. \cite{MP}) we find that the density of variables $\xi_{j}$, $\rho(\xi)$, in the
thermodynamic limit, satisfies the following integral equation
\EQ
\psi^{\prime}(\xi) +2 \pi \rho(\xi) = \int_{-\infty}^{+\infty} \varphi^{\prime}(\xi-u) \rho(u) d u
\EN
where the prime symbol stands for the derivative and the functions $\psi(\xi)$ and 
$\varphi(\xi)$ are given by
\EQ
\psi(\xi) = 2[\arctan(\frac{\xi}{3/2 -b})-\arctan(\frac{\xi}{b+1/2})], ~~
\varphi(\xi) = 2 \arctan(\xi/2)
\EN

This integral equation is solved by Fourier techniques and we find that
\EQ
\rho(\xi) = \frac{\sin[\pi (1/2-b)]}{\cosh(\pi \xi) + \cos[(1/2-b)\pi]}
\EN

Since the $spl(2|1)$ Hamiltonian can be interpreted as spin-$3/2$ chain, the
magnetization per site is then given by
\EQ
\frac{M_{ag}}{L} = \frac{3}{2} - \frac{n+m}{L}
\EN
where the integers $n$ and $m$ (see Bethe ansatz equation) are the number
 of variables 
$\lambda_{j}$ and $\mu_{l}$, respectively. From equation (106) we 
find that $\frac{n}{L} =
2(1/2-b)$ and $\frac{m}{L} = (1/2-b)$, and as consequence of 
equation (107) we
have\footnote{ It is remarkable that although we have considered $b<<1/2 $, the
limit $ b \rightarrow 1/2 $ in equation (108) reproduces the ferromagnetic
structure found for $ b > 1/2 $ .} 
\EQ
\frac{M_{ag}}{L} = 3b
\EN

It is also possible to verify that  the energy per particle associated with
the  structure of zeros (103) is lower than $E^{f_{1}}_{0}(b)$. 
In summary, we believe that 
our results lead to the following picture. Strictly at $b=0$ the system has  zero
magnetization and presents an antiferromagnetic behaviour. As soon as we turn $b>0$, the
ground state gets a finite magnetization and is partially ferromagnetic ordered. We
believe that this picture remains in the whole regime $0<b<1/2$. After the singular
point $b=1/2$($b>1/2$), the system is then fully ordered in the ferromagnetic state
$\ket{\phi^{f_{2}}_{0}}$. In this sense the parameter b plays the role of the 
incommensurability such as a chemical potential or a magnetic field indicating that
at $b=1/2$ the system presents a phase-transition of Pokrovsky-Talapov type \cite{PK}.
In fact, the typical quadratic form of the dispersion relation appears in 
$E^{f_{2}}_{1,2}(b,p)$ for the low-lying excitation around  $p\sim \pi$. It 
seems interesting to understand this picture in terms of the particles and 
anti-particles of a factorizable $S$-matrix. We suspect that the point $b=1/2$ behaves
as the threshold for the mass of the solitons and the 
anti-solitons present in the
theory.

\section{Conclusions}

We have shown that the one parameter family of an integrable $spl(2|1)$ vertex
model is exactly solved by the algebraic Bethe ansatz. The 
eigenstates have been 
formulated in terms of the creation operators 
through the recurrence relation (51).
The eigenvalues of the corresponding transfer matrix is computed by solving an auxiliary problem related to an 
inhomogeneous $6$-vertex model. We have discussed how the 
reflection symmetry $b \rightarrow -b$ can be encoded in terms of the commutation
rules. The ground state picture has been discussed and we 
have presented arguments that the system has a 
commensurable/incommensurable phase transition of Pokrovski-Talapov 
type.

We believe that the formulation described 
in this paper is by no means only
particular to the isotropic $spl(2|1)$ vertex model. As we have 
already commented, the 
whole construction can be generalized for the anisotropic model(trigonometric case)
almost directly. The study of the phase-diagram of the anisotropic model and in 
particular the Pokrovski-Talapov transition, 
seems to us to be a very interesting 
problem. Besides such direct generalization, we have reasons to think that our 
formulation is the cornerstone to 
solve, by the algebraic Bethe ansatz approach,  certain 
integrable models related with the symmetry $C_{n}$. This should be the case of the
isotropic $Sp(2n)$ and $OSP(2|2n-2)$ vertex models \cite{PMSP}. Other model that is still waiting
for an algebraic solution is the Hubbard model. Since this system 
also possesses a hidden $6$-
vertex symmetry, we strongly believe that our formalism can be adapted in order to
give the algebraic Bethe ansatz solution of the Hubbard model \cite{PMH}.
We hope to report on these problems in a future publications.

\section*{Acknowledgements}
We thank F.C. Alcaraz and A. Malvezzi for helpful discussions on the Pokrovski-Talapov 
transition. We also thank F.C. Alcaraz   for pointing us ref. \cite{BS}. The work of M.J. Martins was 
partially supported by Cnpq and FAPESP ( Brazilian agencies). The work of
P.B. Ramos was supported by FAPESP . 

\vspace{0.5cm}

\centerline{\bf Appendix A : The $spl(2|1)$ algebra and the R-matrix properties }
\setcounter{equation}{0}
\renewcommand{\theequation}{A.\arabic{equation}}
The algebra $spl(2|1)$ \cite{RI} consists of four even generators $ \{ S_{\pm}, S_3, B \} $ and
four odd generators $\{ V_{\pm}, \overline{V}_{\pm} \} $ . Following refs. \cite{ITO,MA} they
satisfy the commutation rules
\EQ
[S_3, S_{\pm}] = \pm S_{\pm},~~ [S_{+},S_{-}]=2 S_3,~~ 
\{ V_{\pm},\overline{V}_{\pm} \} = \pm
1/2 S_{\pm}
\EN
\EQ
\{ V_{+}, \overline{V}_{-} \} =-1/2 P_{+},~~ \{ \overline{V}_{+}, V_{-} \} = -1/2 P_{-},~~
[P_{\pm},\overline{V}_{\pm}] = \pm \overline{V}_{\pm},~~[P_{\mp},V_{\pm}]= \pm V_{\pm}
\EN
\EQ
[P_{\pm},V_{\pm}]=[P_{\mp},\overline{V}_{\pm}]=0; \{ V_i,V_j \}= \{\overline{V}_i, \overline{V}_j \} =0, i,j= \pm 
\EN
where the symbols $[,]$ and $ \{ , \} $ denote the comutator and the anti-commutator, respectively . We also have the identity $ P_{\pm} = S_3 \pm B $ . The 4-dimensional representation
\cite{RI,MA} of these generators in the  $bffb$ grading  possesses the following
matrix representations 
\EQ
V_{+} =\left( \begin{array}{cccc} 
	0 & \epsilon & 0& 0 \\
	0 & 0 & 0 & 0 \\
	0 & 0 & 0 & \alpha \\
	0 & 0 & 0 & 0\\
	\end{array}
	\right) ~~ 
V_{-} =\left( \begin{array}{cccc} 
	0 & 0 & 0 & 0 \\
	0 & 0 & 0 & 0 \\
	-\alpha & 0 & 0 & 0 \\
	0 & \epsilon & 0 & 0\\
	\end{array}
	\right) ~~ 
P_{+} =\left( \begin{array}{cccc} 
	1/2-b & 0 & 0 & 0\\
	0 & 1/2-b & 0  & 0\\
	0 & 0 & -1/2-b  & 0\\
	0 & 0 & 0 & -1/2-b \\
	\end{array}
	\right) 
\EN
\EQ
\overline{V}_{+} =\left( \begin{array}{cccc} 
	0 &  0 & \gamma & 0 \\
	0 & 0 & 0 & \beta \\
	0 & 0 & 0 & 0 \\
	0 & 0 & 0 & 0\\
	\end{array}
	\right) ~~
\overline{V}_{-} =\left( \begin{array}{cccc} 
	0 & 0 & 0 & 0 \\
	-\beta & 0 & 0 & 0 \\
	0 & 0 & 0 & 0 \\
	0 & 0 & \gamma & 0\\
	\end{array}
	\right) ~~ 
P_{-} =\left( \begin{array}{cccc} 
	1/2+b & 0 & 0 & 0\\
	0 & -1/2+b & 0  & 0\\
	0 & 0 & 1/2+b  & 0\\
	0 & 0 & 0 & -1/2+b \\
	\end{array}
	\right) 
\EN
\EQ
S_{+} =\left( \begin{array}{cccc} 
	0 & 0 & 0 & 1\\
	0 & 0 & 0  & 0\\
	0 & 0 & 0  & 0\\
	0 & 0 & 0 & 0 \\
	\end{array}
	\right) ~~
S_{-} =\left( \begin{array}{cccc} 
	0 & 0 & 0 & 0\\
	0 & 0 & 0  & 0\\
	0 & 0 & 0  & 0\\
	1 & 0 & 0 & 0 \\
	\end{array}
	\right) ~~
S_{3} =\left( \begin{array}{cccc} 
	1/2 & 0 & 0 & 0\\
	0 & 0 & 0  & 0\\
	0 & 0 & 0  & 0\\
	0 & 0 & 0 & -1/2 \\
	\end{array}
	\right)
\EN
where $4 \alpha \gamma = 1+2b $ and $4 \beta \epsilon =1 -2b $ . The Casimir operator is
written in terms of these generators as
\bear
C_{i,i+1}(b) = 2 \{ S_+ \stackrel{s}{\otimes} S_{-} +
 S_{-} \stackrel{s}{\otimes} S_{+} \} + 
 2 \{ P_{+} \stackrel{s}{\otimes} P_{-}  + 
  P_{+} \stackrel{s}{\otimes} P_{-} \}  + 
\nonumber \\
4 \{ V_{-} \stackrel{s}{\otimes} \overline{V}_{+} +
 \overline{V}_{-} \stackrel{s}{\otimes} V_{+} -
 \overline{V}_{+} \stackrel{s}{\otimes} V_{-} -
 V_{+} \stackrel{s}{\otimes} \overline{V}_{-} \} +4b^2 I
\ear

In equation (A.7) the  symbol $\stackrel{s}{\otimes}$ stands for
the supertensor product. More 
precisely the elements of $ A \stackrel{s}{\otimes} B $ are
\EQ
(A \stackrel{s}{\otimes} B)_{ab}^{ij} = (-1)^{p(i)p(j) +p(a)p(b) +p(i)p(B)} A_{ai} B_{bj}
\EN
where $p(f)$ is the Grassmann parity of the object $f$ .

As it has been observed by Maassarani \cite {MA} the projectors $P_i, i=1,2,3 $ of the 
$spl(2|1) $ algebra play a fundamental role in the construction of the $R$-matrix . In the
isotropic limit, following ref. \cite{MA}, we find that the $ 16 \times 16 $ matrices expressions
for the projectors $P_i$ in the $fbbf$ grading are
\EQ
{\small
 P_1(b) =
   \pmatrix{
 1 &0 &0 &0 &0  &0  &0  &0  &0  &0  &0  &0  &0  &0  &0  &0    \cr
 0 & 1/2 &0 &0 &-1/2  &0  &0  &0  &0  &0  &0  &0  &0  &0  &0  &0    \cr
 0 &0 &1/2 &0 &0  &0  &0  &0  &-1/2  &0  &0  &0  &0  &0  &0  &0    \cr
 0 &0 &0 &\tilde{\alpha}(b) &0  &0  &\tilde{\gamma}(b) &0  &0  &-\tilde{\alpha}(b)  
&0  &0  &\tilde{\gamma}(b)  &0  &0  &0    \cr
 0 &-1/2 &0 &0 &1/2  &0  &0  &0  &0  &0  &0  &0  &0  &0  &0  &0    \cr
 0 &0 &0 &0 &0  &0  &0  &0  &0  &0  &0  &0  &0  &0  &0  &0    \cr
 0 &0 &0 &-\tilde{\gamma}(b) &0  &0  &\tilde{\beta}(b) &0  &0  &-\tilde{\beta}(b)  
&0  &0  &-\tilde{\gamma}(b)  &0  &0  &0    \cr
 0 &0 &0 &0 &0  &0  &0  &0  &0  &0  &0  &0  &0  &0  &0  &0    \cr
 0 &0 &-1/2 &0 &0  &0  &0  &0  &1/2  &0  &0  &0  &0  &0  &0  &0    \cr
 0 &0 &0 &\tilde{\gamma}(b) &0  &0  &-\tilde{\beta}(b)  &0  &0  &\tilde{\beta}(b)  
&0  &0  &\tilde{\gamma}(b)  &0  &0  &0    \cr
 0 &0 &0 &0 &0  &0  &0  &0  &0  &0  &0  &0  &0  &0  &0  &0    \cr
 0 &0 &0 &0 &0  &0  &0  &0  &0  &0  &0  &0  &0  &0  &0  &0    \cr
 0 &0 &0 &\tilde{\alpha}(b) &0  &0  &\tilde{\gamma}(b)  &0  &0  &-\tilde{\gamma}(b)  
&0  &0  &\tilde{\alpha}(b)  &0  &0  &0    \cr
 0 &0 &0 &0 &0  &0  &0  &0  &0  &0  &0  &0  &0  &0  &0  &0    \cr
 0 &0 &0 &0 &0  &0  &0  &0  &0  &0  &0  &0  &0  &0  &0  &0    \cr
 0 &0 &0 &0 &0  &0  &0  &0  &0  &0  &0  &0  &0  &0  &0  &0    \cr}
\nonumber \\
}
\EN
\EQ
{\small
 P_3(b) =
   \pmatrix{
 0 &0 &0 &0 &0  &0  &0  &0  &0  &0  &0  &0  &0  &0  &0  &0    \cr
 0 & 0 &0 &0 &0  &0  &0  &0  &0  &0  &0  &0  &0  &0  &0  &0    \cr
 0 &0 &0 &0 &0  &0  &0  &0  &0  &0  &0  &0  &0  &0  &0  &0    \cr
 0 &0 &0 &\tilde{\beta}(b) &0  &0  &-\tilde{\gamma}(b) &0  &0  &\tilde{\gamma}(b)  
&0  &0  &\tilde{\beta}(b)  &0  &0  &0    \cr
 0 &0 &0 &0 &0  &0  &0  &0  &0  &0  &0  &0  &0  &0  &0  &0    \cr
 0 &0 &0 &0 &0  &0  &0  &0  &0  &0  &0  &0  &0  &0  &0  &0    \cr
 0 &0 &0 &\tilde{\gamma}(b) &0  &0  &\tilde{\alpha}(b) &0  &0  &-\tilde{\alpha}(b)  
&0  &0  &\tilde{\gamma}(b)  &0  &0  &0    \cr
 0 &0 &0 &0 &0  &0  &0  &1/2  &0  &0  &0  &0  &0  &-1/2  &0  &0    \cr
 0 &0 &0 &0 &0  &0  &0  &0  &0  &0  &0  &0  &0  &0  &0  &0    \cr
 0 &0 &0 &-\tilde{\gamma}(b) &0  &0  &-\tilde{\alpha}(b)  &0  &0  &\tilde{\alpha}(b)  
&0  &0  &-\tilde{\gamma}(b)  &0  &0  &0    \cr
 0 &0 &0 &0 &0  &0  &0  &0  &0  &0  &0  &0  &0  &0  &0  &0    \cr
 0 &0 &0 &0 &0  &0  &0  &0  &0  &0  &0  &1/2  &  &0  &-1/2  &0    \cr
 0 &0 &0 &\tilde{\beta}(b) &0  &0  &-\tilde{\gamma}(b)  &0  &0  &\tilde{\gamma}(b)  
&0  &0  &\tilde{\beta}(b)  &0  &0  &0    \cr
 0 &0 &0 &0 &0  &0  &0  &-1/2  &0  &0  &0  &0  &0  &1/2  &0  &0    \cr
 0 &0 &0 &0 &0  &0  &0  &0  &0  &0  &0  &-1/2  &0  &0  &1/2  &0    \cr
 0 &0 &0 &0 &0  &0  &0  &0  &0  &0  &0  &0  &0  &0  &0  &1    \cr}
\nonumber \\
}
\EN
where $ P_2(b) = I - (P_1(b) +P_3(b)) 
$ \cite{MA} and $ \tilde{\alpha}(b)=(2b+1)/8b $, 
$\tilde{\beta}(b) = (2b-1)/8b $ and 
$ \tilde{\gamma}(b) = \sqrt{1-4b^2}/8b $ . Besides the
usual projectors identities, $ P_i(b) P_j(b) = \delta_{i,j} P_j(b) $, 
{the operators $P_1(b)$ and $P_3(b)$
satisfy the following useful relations
\EQ
P^g = I -2[P_1(b) +P_3(b)] ,~~ P^g P_1(b) = -P_1(b),~~ P^g P_3(b) = -P_3(b) 
\EN
where $(P^g)_{ij}^{lk}= (-1)^{p(i)p(j)} \delta_{i,k}\delta_{j,l}$ 
defines the graded  
permutation operator. By using definitions (A.7,A.8) and 
expressions (A.9,A.10), the Casimir invariant
can be connected to the projectors $P_1(b)$ and $P_3(b)$ as
\EQ
C_{i,i+1}(b) = I -2[P_1(b) +P_3(b)] + 4b [P_1(b) -P_3(b)]
\EN

Such relations are important in the study of the classical limit of the $spl(2|1)$ $R$-matrix .
In fact, by introducing the quasi-classical parameter $\eta$ in equation (1) as $\lambda
\rightarrow \frac{\lambda}{\eta} $ we have
\EQ
R(\lambda,b,\eta)= I -\frac{4 \lambda}{(1-2b) \eta +2 \lambda} P_1(b) 
 -\frac{4 \lambda}{(1+2b) \eta +2 \lambda} P_3(b) 
\EN

By expanding this last expression around $\eta=0$ we find
\EQ
R(\lambda,b,\eta=0) = I -2[P_1(b) +P_3(b)]  =P^g
\EN
and
\EQ
\frac{\partial R(\lambda,b,\eta)}{\partial \eta}|_{\eta=0} = \frac{1}{\lambda}
\left \{ P_1(b) +P_3(b)
 -2b[P_1(b)-P_3(b)] \right \}
\EN

As a consequence, we find that  equation (4) of section 2 
follows  from the  expressions
(A.14,A.15) and the identity (A.12). Analogously, the corresponding Hamiltonian can be also written in terms of the
Casimir operator . Considering that the 
two-body Hamiltonian $H_{i,i+1} $ is determined as the
derivative of the $R$-matrix at the regular point $ \lambda =0 $ , we obtain
\EQ
H_{i,i+1} = -\frac{4}{1-2b}P_1(b) -\frac{4}{1+2b}P_3(b)
\EN
 
Now if we consider the identities (A.11,A.12) the square of the Casimir operator is
\EQ
C_{i,i+1}^2(b) = I +8b[P_3(b)-P_1(b)] +16 b^2[P_3(b)+P_1(b)]
\EN
and now by solving equations (A.12) and 
(A.17) for the operators $P_1(b)$ and $P_3(b)$ we find
\EQ
\frac{P_1(b)}{1+2b} =\frac{1}{16b(1-2b)(1+2b)}[ (4b+1)I -4b C_{i,i+1}(b) -C_{i,i+1}^2(b) ]
\EN
\EQ
\frac{P_3(b)}{1-2b} =\frac{1}{16b(1-2b)(1+2b)}[ (4b-1)I -4b C_{i,i+1}(b) +C_{i,i+1}^2(b) ]
\EN

These last relations are then used in equation (A.16) in order to
reproduce 
the Hamiltonian
expression (10) presented in section 3.

Finally, we remark that the $spl(2|1) $ algebra is invariant 
under the following isomorphic transformation
\EQ
S_{\pm} \rightarrow S_{\pm},~~ S_{3} \rightarrow S_{3},~~ P_{\pm} \rightarrow
P_{\mp},~~ V_{\pm} \rightarrow \overline{V}_{\pm},~~
 \overline{V}_{\pm} \rightarrow V_{\pm}
\EN
which in terms of the parameter $b$ means the reflection symmetry $b \rightarrow-b $. This can be easily seen from the matrix representations 
(A.4,A.5,A.6) of the $spl(2|1)$ generators, provided we 
also perform the canonical transformation 
$f_1 \leftrightarrow f_2 $.
\vspace{0.5cm}

\centerline{\bf Appendix B : The two-particle state }
\setcounter{equation}{0}
\renewcommand{\theequation}{B.\arabic{equation}}

The main purpose of this Appendix is 
to give extra details that the two-particle state (45) 
we have constructed is in fact an eigenstate under 
the Bethe ansatz restriction (42). In order to 
collect the unwanted terms we have to use the annihilation property (16) 
and the following additional
commutation rules
\EQ
C_{a}(\lambda)B_{b}(\mu) = 
B_{b}(u)C_{a}(\lambda) - \frac{m(\lambda-\mu)}
{p(\lambda-\mu)}[B(\lambda)A_{ab}(\mu) - B(\mu)A_{ab}(\lambda)],~~a=1,2
\EN
\bear
C_{a+3}(\lambda)B_{b}(\mu) = 
\frac{s(\lambda-\mu)}{p(\lambda-\mu)}B_{a}(\mu)C_{b+3}(\lambda) +
 \frac{t(\lambda-\mu)}{p(\lambda-\mu)}B_{b}(u)C_{a+3}(\lambda)
\nonumber \\
- \xi_{ab}\frac{\sigma(\lambda-\mu)}{p(\lambda-\mu)} \{ F(\mu)C_{3}(\lambda) + 
B(\mu)D(\lambda) \}
- \frac{n(\lambda-\mu)}{p(\lambda-\mu)}B_{a}(\lambda)C_{b+3}(\mu) 
\nonumber \\
+ \frac{\sigma(\lambda-\mu)}{p(\lambda-\mu)} 
\xi_{lm} A_{la}(\lambda)A_{mb}(\mu),~~a=b=1,2
\ear
\bear
E_{a}(\lambda)B_{b}(\mu) =
\frac{g(\lambda-\mu)}{f(\lambda-
\mu)}B_{b}(\mu)E_{a}(\lambda) - 
\frac{m(\lambda-\mu)}{f(\lambda-\mu)}F(\lambda)A_{ab}(\mu)  
\nonumber \\
+\frac{q(\lambda-\mu)}{f(\lambda-
\mu)}F(\mu)A_{ab}(\lambda),~~ a=b=1,2
\ear

By combining such commutation rules together with those mentioned in
section 4 and property (15) we find that the non-trivial unwanted terms have
the following structures
\EQ
B_{a}(\lambda)B_{b}(\lambda_{i}); ~~ E_{a}(\lambda)B_{b}(\lambda_{i}); ~~ F(\lambda)(F^{12}-F^{21})
\EN

For the first two cases in equation (B.4) it is enough to fix 
$\lambda_{i}=\lambda_{1}$, since the other possibility ($\lambda_{i}=\lambda_{2}$)
 has already been discussed in the main text (see section 5). We now
summarize the functional forms which multiply the terms (B.4), 
and discuss how
they are canceled out.

$1.$ The $B_{a}(\lambda)B_{a}(\lambda_{1})$ term, 
for a given index $a$, appears as
\bear
\left \{
\frac{m(\lambda_{1}-\lambda)}{f(\lambda_{1}-\lambda)}\frac{m(\lambda_{2}-\lambda_{1})}{f(\lambda_{2}-\lambda_{1})} -
\frac{m(\lambda_{2}-\lambda)}{f(\lambda_{2}-\lambda)}\frac{r_{aa}^{aa}(\lambda_{1}-\lambda)}{f(\lambda_{1}-\lambda)}
\right \}F^{aa}[l(\lambda_{2})]^L +
\nonumber \\
\left \{
\frac{m(\lambda-\lambda_{1})}{f(\lambda-\lambda_{1})}\frac{m(\lambda_{1}-\lambda_{2})}{f(\lambda_{1}-\lambda_{2})} -
\frac{m(\lambda-\lambda_{2})}{f(\lambda-\lambda_{2})}\frac{r_{aa}^{aa}(\lambda-\lambda_{1})r_{aa}^{aa}(\lambda-\lambda_{2})}
{f(\lambda-\lambda_{1})l(\lambda_{1}-\lambda)}
\right \}F^{aa}[f(\lambda_{2})]^L
\ear
and by using the identities (36) and (41), we find
\EQ
\left \{ -\frac{m(\lambda_2-\lambda)}{f(\lambda_2 -\lambda) f(\lambda_1- \lambda)}+ \frac{m(\lambda_{1}-\lambda)}{f(\lambda_{1}-\lambda)}
\frac{m(\lambda_{2}-\lambda_{1})}{f(\lambda_{2}-\lambda_{1})}
\{ [l(\lambda_{2})]^L + [f(\lambda_{2})]^L \} \right \} F^{aa}
\EN
which is null by taking $a=b$ in the Bethe ansatz equation (42).

$2.$ The $B_{a}(\lambda)B_{b}(\lambda_{1})$ term for $a \neq b$ 
is more involved. Here the cases $B_{1}(\lambda)B_{2}(\lambda_{2})$ and $B_{2}(\lambda)B_{1}(\lambda_{2})$  are equivalent. The functional structure which 
comes from the creation operator $F(\lambda_{1})$ of the two-particle 
eigenstate is
\EQ
\frac{\sigma(\lambda_{1}-\lambda)}{p(\lambda_{1}-\lambda)}\frac{\sigma(\lambda_{1}-\lambda_{2})}{p(\lambda_{1}-\lambda_{2})}
[l(\lambda_{2})]^L (F^{12}-F^{21})
\EN
and that coming from $B_{a}(\lambda_{1})B_{b}(\lambda_{2})$ are
\bear
 \left \{
\frac{m(\lambda_{1}-\lambda)}{f(\lambda_{1}-\lambda)}\frac{m(\lambda_{2}-\lambda_{1})}{f(\lambda_{2}-\lambda_{1})}F^{21} -
\frac{m(\lambda_{2}-\lambda)}{f(\lambda_{2}-\lambda)}
\frac{[r_{12}^{12}(\lambda_{1}-\lambda)F^{21}
+r_{12}^{21}(\lambda_{1}-\lambda)F^{12}]}{f(\lambda_{1}-\lambda)}
\right \}[l(\lambda_{2})]^L +
\nonumber \\
\left \{
\frac{m(\lambda-\lambda_{1})}{f(\lambda-\lambda_{1})}\frac{m(\lambda_{1}-\lambda_{2})}{f(\lambda_{1}-\lambda_{2})} -
\frac{m(\lambda-\lambda_{2})}{f(\lambda-\lambda_{2})}\frac{1}
{f(\lambda-\lambda_{1})l(\lambda_{1}-\lambda)}
\right \}F^{12}[f(\lambda_{2})]^L \nonumber \\
\ear

The simplest way to see that this term is null is to use 
the Bethe ansatz equations (42) for $\lambda_{i}=\lambda_{2}$
in order to have only terms proportional 
to $F^{21}[f(\lambda_{2})]^L$ and 
$F^{12}[f(\lambda_{2})]^L$. In particular we have
\EQ
[l(\lambda_{2})]^L (F^{12}-F^{21}) = 
-[f(\lambda_{2})]^L \left \{ r_{12}^{12}(\lambda_{2}-\lambda_{1}) 
- r_{12}^{12}(\lambda_{2}-\lambda_{1}) \right \} (F^{21}-F^{12})
\EN

By making such manipulations and collecting 
the term proportional to $F^{21}[f(\lambda_{2})]^L$ of expressions 
(B.7) and (B.8) we have 
\bear
\left \{
-\frac{m(\lambda_{1}-\lambda)}{f(\lambda_{1}-\lambda)}\frac{m(\lambda_{2}-\lambda_{1})}{f(\lambda_{2}-\lambda_{1})}
r_{21}^{12}(\lambda_{2}-\lambda_{1}) -
\frac{\sigma(\lambda_{1}-\lambda)}{p(\lambda_{1}-\lambda)}\frac{\sigma(\lambda_{1}-\lambda_{2})}{p(\lambda_{1}-\lambda_{2})}
[r_{12}^{12}(\lambda_{2}-\lambda_{1}) - r_{12}^{21}(\lambda_{2}-\lambda_{1})]
\right . \nonumber \\ \left .
+\frac{m(\lambda_{2}-\lambda)}{f(\lambda_{2}-\lambda)}\frac{1}{f(\lambda_{1}-\lambda)}
[r_{12}^{12}(\lambda_{1}-\lambda)r_{21}^{12}(\lambda_{2}-\lambda_{1}) + 
r_{12}^{21}(\lambda_{1}-\lambda)r_{12}^{12}(\lambda_{2}-\lambda_{1})] 
\right \}F^{21}[f(\lambda_{2})]^L \nonumber \\
\ear
which is in fact null if one uses the identity (47) and 
the factorization relation
\bear
-\frac{\sigma(\lambda_{1}-\lambda)}{p(\lambda_{1}-\lambda)}\frac{\sigma(\lambda_{2}-\lambda_{1})}{p(\lambda_{2}-\lambda_{1})}
(\lambda_{2}-\lambda_{1} +1) =
\frac{m(\lambda_{1}-\lambda)}{f(\lambda_{1}-\lambda)}
\frac{m(\lambda_{2}-\lambda_{1})}{f(\lambda_{2}-\lambda_{1})}
(\lambda_{2}-\lambda_{1}) 
\nonumber \\
-\frac{m(\lambda_{2}-\lambda)}{f(\lambda_{2}-\lambda)}\frac{1}{f(\lambda_{1}-\lambda)}
\frac{(\lambda_{2}-\lambda)}{(\lambda_{1}-\lambda + 1)} \nonumber \\
\ear

Analogously, the terms proportional to $F^{12}[f(\lambda_{2})]^L$ are
\bear
\left \{
-\frac{m(\lambda_{1}-\lambda)}{f(\lambda_{1}-\lambda)}\frac{m(\lambda_{2}-\lambda_{1})}{f(\lambda_{2}-\lambda_{1})}
r_{21}^{21}(\lambda_{2}-\lambda_{1}) + 
\frac{m(\lambda-\lambda_1)}{f(\lambda-\lambda_1)}
\frac{m(\lambda_1-\lambda_2)}{f(\lambda_1-\lambda_2)} +
\frac{m(\lambda_{2}-\lambda)}{f(\lambda_{2}-\lambda)}
\frac{1}{f(\lambda_{1}-\lambda)}
\right . \nonumber \\ \left .
[r_{12}^{12}(\lambda_{1}-\lambda)r_{21}^{21}(\lambda_{2}-\lambda_{1}) + 
r_{12}^{21}(\lambda_{1}-\lambda)r_{12}^{21}(\lambda_{2}-\lambda_{1})] -
\frac{m(\lambda-\lambda_{2})}{f(\lambda-\lambda_{2})}\frac{1}
{f(\lambda-\lambda_{1})l(\lambda_{1}-\lambda)}
\right . \nonumber \\ \left .
+\frac{\sigma(\lambda_{1}-\lambda)}{p(\lambda_{1}-\lambda)}\frac{\sigma(\lambda_{1}-\lambda_{2})}{p(\lambda_{1}-\lambda_{2})}
[r_{12}^{12}(\lambda_{1}-\lambda) - r_{21}^{12}(\lambda_{2}-\lambda_{1})]
\right \}F^{12}[f(\lambda_{2})]^L
\ear

 Now, by noticing that the first two terms in (B.12) can be simplified as
\EQ
\frac{m(\lambda_{1}-\lambda)}{f(\lambda_{1}-\lambda)}\frac{m(\lambda_{2}-\lambda_{1})}{f(\lambda_{2}-\lambda_{1})}
r_{12}^{21}(\lambda_{2}-\lambda_{1})
\EN
one can proceed as we did before. If we use the simplification (B.11)
and the last relation (B.13) we finally find that the term
proportional to $F^{12}[f(\lambda_{2})]^L $ is also null.

$3$. The $E_{a}(\lambda)B_{b}(\lambda_{2})$ term for $a \neq b$ appears as
\bear
\left \{
\frac{\sigma(\lambda-\lambda_{2})}{p(\lambda-\lambda_{2})}\frac{f(\lambda-\lambda_{1})}{p(\lambda-\lambda_{1})} -
\frac{\sigma(\lambda-\lambda_{1})}{p(\lambda-\lambda_{1})}\frac{m(\lambda_{1}-\lambda_{2})}{f(\lambda_{1}-\lambda_{2})}
\right \}F^{aa}[f(\lambda_{2})]^L +
\nonumber \\
\left \{
\frac{\sigma(\lambda-\lambda_{1})}{p(\lambda-\lambda_{1})}\frac{m(\lambda_{2}-\lambda_{1})}{f(\lambda_{2}-\lambda_{1})} -
\frac{\sigma(\lambda-\lambda_{2})}{p(\lambda-\lambda_{2})}\frac{r_{21}^{12}(\lambda-\lambda_{1})}
{g(\lambda-\lambda_{1})}
\right \}F^{aa}[l(\lambda_{2})]^L
\ear
and by using the identity
\EQ
\frac{f(\lambda)}{p(\lambda)} = - \frac{r_{21}^{12}(\lambda)}{g(\lambda)}
\EN
we are able to  simplify equation (B.14) as
\bear
\left \{
\frac{\sigma(\lambda-\lambda_{2})}{p(\lambda-\lambda_{2})}\frac{f(\lambda-\lambda_{1})}{p(\lambda-\lambda_{1})} +
\frac{\sigma(\lambda-\lambda_{1})}{p(\lambda-\lambda_{1})}\frac{m(\lambda_{2}-\lambda_{1})}{f(\lambda_{2}-\lambda_{1})}
\right \}F^{aa} \left ( [l(\lambda_{2})]^L +[f(\lambda_{2})]^L \right )
\ear
which is automatically null by taking $a=b$ in the Bethe ansatz equations (42).

$4$. The $E_{a}(\lambda)B_{a}(\lambda_{2})$ term has a contribution 
from both creation operators: $B_{i}(\lambda_{1})B_{j}(\lambda_{2})$ 
and $F(\lambda_{1})$. The contribution coming from $F(\lambda_{1})$ is
\EQ
-\frac{q(\lambda-\lambda_{1})}{g(\lambda-\lambda_{1})}\frac{\sigma(\lambda_{1}-\lambda_{2})}{p(\lambda_{1}-\lambda_{2})}
[l(\lambda_{2})]^L (F^{12}-F^{21})
\EN
and those comming from $B_{i}(\lambda_{1})B_{j}(\lambda_{2})$  are
\bear
 \left \{
\frac{\sigma(\lambda-\lambda_{1})}{p(\lambda-\lambda_{1})}\frac{m(\lambda_{2}-\lambda_{1})}{f(\lambda_{2}-\lambda_{1})}F^{12} -
\frac{\sigma(\lambda-\lambda_{2})}{p(\lambda-\lambda_{2})}\frac{[r_{11}^{11}(\lambda-\lambda_{1})F^{21}
-r_{12}^{12}(\lambda-\lambda_{1})F^{12}]}{g(\lambda-\lambda_{1})}
\right \}[l(\lambda_{2})]^L +
\nonumber \\
\left \{
\frac{\sigma(\lambda-\lambda_{2})}{p(\lambda-\lambda_{2})}\frac{f(\lambda-\lambda_{1})}{p(\lambda-\lambda_{1})} -
\frac{\sigma(\lambda-\lambda_{1})}{p(\lambda-\lambda_{1})}\frac{m(\lambda_{1}-\lambda_{2})}
{f(\lambda_{1}-\lambda_{2})}
\right \}F^{21}[f(\lambda_{2})]^L \nonumber \\
\ear

In order to show that all these terms together are in fact 
null, one can follow the same steps of the 
procedure which we have used for the term $B_{i}(\lambda_{1})
B_{j}(\lambda_{2})$  ($i \neq j$). However, the crucial factorization
relation here is a bit different, namely
\bear
\frac{\sigma(\lambda-\lambda_{1})}{p(\lambda-\lambda_{1})}\frac{m(\lambda_{2}-\lambda_{1})}{f(\lambda_{2}-\lambda_{1})}
r_{12}^{21}(\lambda_{2}-\lambda_{1}) =
-\frac{q(\lambda-\lambda_{1})}{g(\lambda-\lambda_{1})}\frac{\sigma(\lambda_{2}-\lambda_{1})}{p(\lambda_{2}-\lambda_{1})}
\nonumber \\
+\frac{\sigma(\lambda-\lambda_{2})}{p(\lambda-\lambda_{2})}\frac{1}{g(\lambda-\lambda_{1})}\frac{(\lambda-\lambda_{2}+1)}
{(\lambda_{2}-\lambda_{1} + 1)(\lambda-\lambda_{1} + 1)} \nonumber \\
\ear

$5$. To collect all the non-trivial unwanted terms proportional to $F(\lambda)(F^{12}-F^{21})$ is a very cumbersome job. The main
reason is that all the diagonal 
operators $\sum_{a=1}^{2} A_{aa}(\lambda)$,$B(\lambda)$ and $D(\lambda)$ give non-trivial contributions
which are proportional to the many combinations of $[l(\lambda_{1})]^L
\{(\cdots)[f(\lambda_{2})]^L+(\cdots)[l(\lambda_{2})]^L \}$ and 
$[f(\lambda_{1})]^L \{(\cdots)[f(\lambda_{2})]^L+(\cdots)
[l(\lambda_{2})]^L \}$. For 
instance, the terms proportional to $[l(\lambda_{1})]^L$ 
are
\bear
\left \{
-\frac{\sigma(\lambda-\lambda_{1})}{p(\lambda-\lambda_{1})}\frac{m(\lambda-\lambda_{2})}{f(\lambda-\lambda_{2})}
\frac{1}{f(\lambda-\lambda_{1})l(\lambda_{1}-\lambda)} +
\frac{\sigma(\lambda-\lambda_{1})}{p(\lambda-\lambda_{1})}\frac{m(\lambda-\lambda_{1})}{f(\lambda-\lambda_{1})}
\frac{m(\lambda_{1}-\lambda_{2})}{f(\lambda_{1}-\lambda_{2})}
\right \}[f(\lambda_{2})]^L +
\nonumber \\
\left \{
\frac{\sigma(\lambda_{1}-\lambda_{2})}{p(\lambda_{1}-\lambda_{2})}\frac{n(\lambda_{1}-\lambda)}{p(\lambda_{1}-\lambda)} +
\frac{\sigma(\lambda_{1}-\lambda)}{p(\lambda_{1}-\lambda)}\frac{m(\lambda_{2}-\lambda)}
{f(\lambda_{2}-\lambda)f(\lambda_{1}-\lambda)}
\right \}[l(\lambda_{2})]^L \nonumber \\
\ear
while  those proportional to $[f(\lambda_{1})]^L$ are
\bear
\left \{
-\frac{\sigma(\lambda-\lambda_{2})}{p(\lambda-\lambda_{2})}\frac{m(\lambda-\lambda_{1})}{p(\lambda-\lambda_{1})} +
\frac{\sigma(\lambda_{1}-\lambda_{2})}{p(\lambda_{1}-\lambda_{2})}\frac{n(\lambda-\lambda_{1})}{p(\lambda-\lambda_{1})}
\right \}[f(\lambda_{2})]^L +
\nonumber \\
\left \{
\frac{m(\lambda-\lambda_{1})}{f(\lambda-\lambda_{1})}\frac{1}{g(\lambda-\lambda_{1})}
\frac{\sigma(\lambda-\lambda_{2})}{p(\lambda-\lambda_{2})}[1+r_{21}^{21}(\lambda-\lambda_{1})] -
\right . \nonumber \\ \left .
\frac{\sigma(\lambda-\lambda_{1})}{p(\lambda-\lambda_{1})}\frac{m(\lambda-\lambda_{1})}
{p(\lambda-\lambda_{1})}\frac{m(\lambda_{1}-\lambda_{2})}{f(\lambda_{1}-\lambda_{2})} - 
2\frac{\sigma(\lambda_{1}-\lambda_{2})}{p(\lambda_{1}-\lambda_{2})}\frac{m(\lambda-\lambda_{1})}
{g(\lambda-\lambda_{1})}\frac{q(\lambda-\lambda_{1})}{f(\lambda-\lambda_{1})}
\right \}[l(\lambda_{2})]^L
\ear

In order to cancel these expressions one has to use 
the Bethe ansatz identity (42) for both $\lambda_{1}$ and $\lambda_{2}$.
For instance, we first use such relation for $\lambda_{2}$ in equations (B.20,B.21), and as a result we only have terms proportional to
$[f(\lambda_{2})]^L \{(\cdots)[f(\lambda_{1})]^L+(\cdots)[l(\lambda_{1})]^L \}$. Now, we perform the same operation for $\lambda_{1}$ in
order to eliminate the $[l(\lambda_{1})]^L$ terms. The final expression is very complicated, but we have checked that it is null
with the helping of the $Mathematica^{TM}$ software.

	This finishes our analysis concerning the unwanted terms, and the results of this Appendix together with those of section
5 show that the Bethe ansatz equations (42) 
are sufficient conditions to cancel all of them out.

For sake of completeness let us also discuss the wanted terms. They are 
responsible for the eigenvalues of the transfer matrix .
The wanted terms are constituted by two kinds of creation 
operators and from the two-particle eigenstate expression (45) we have
\EQ
B_a(\lambda_1) B_b(\lambda_2) F^{ba}, ~~ l(\lambda_2)^L \frac{\sigma(\lambda_1- \lambda_2)}{p(\lambda_1-\lambda_2)} F(\lambda_1)
\EN

The contribution proportional to the first term of equation (B.22) can be 
obtained directly from the commutation relations (22,25-27),
and it is easy to get
\EQ
\Lambda(\lambda,\{ \lambda_i \} ) = l(\lambda)^L \prod_{i=1}^{2} \frac{l(\lambda_i-\lambda)}{f(\lambda_i - \lambda)} +
 f(\lambda)^L \prod_{i=1}^{2} \frac{\Lambda^{(1)}(\lambda,\{ \lambda_i \})}{f(\lambda - \lambda_i)} +
p(\lambda)^L \prod_{i=1}^{2} \frac{g(\lambda-\lambda_i)}{p(\lambda - \lambda_i)}
\EN
where $ \Lambda^{(1)}(\lambda,\{ \lambda_i \} ) $ 
is the eigenvalue of the inhomogeneous $6$-vertex system 
with two sites (see equation (61) of section 5 )
. Of course the contribution coming from the second 
term has to be precisely the same as that we have got in
equation (B.23) . Such calculation is a bit more elaborated , since  the action of the diagonal
operators $\sum_{a} A_{aa}(\lambda) $ , $B(\lambda)$ and $D(\lambda) $ on the term $B_a(\lambda_1) B_b(\lambda_2) F^{ba} $
can produce many terms of type $F(\lambda_1) (F^{12}-F^{21}) $ . The analysis is as follows . The terms coming from $D(\lambda)$
are
\EQ
p(\lambda)^L l(\lambda)^L (F^{12}-F^{21}) \left [ \frac{q(\lambda-\lambda_1)}{p(\lambda-\lambda_1)}
\frac{\sigma(\lambda-\lambda_2)}{p(\lambda-\lambda_2)}
-\frac{r(\lambda-\lambda_1)}{p(\lambda-\lambda_1)}
\frac{\sigma(\lambda_1-\lambda_2)}{p(\lambda_1-\lambda_2)} \right ]
\EN
and if we use the factorization identity
\EQ
-\frac{q(\lambda-\lambda_1) \sigma(\lambda-\lambda_2) p(\lambda_1-\lambda_2)}{\sigma(\lambda_1-\lambda_2)} +
r(\lambda-\lambda_1)p(\lambda-\lambda_2) = \prod_{i=1}^2 g(\lambda-\lambda_i)
\EN
we find that the operator $D(\lambda)$ contributes to the eigenvalue with
\EQ
p(\lambda)^L \prod_{i=1}^{2} \frac{g(\lambda-\lambda_i)}{p(\lambda-\lambda_i)}
\EN

Those coming form $B(\lambda) $ are
\EQ
l(\lambda)^L l(\lambda_2)^L (F^{12}-F^{21}) \left [ -\frac{\sigma(\lambda_1-\lambda)
l(\lambda_1-\lambda) m(\lambda_2-\lambda)}{p(\lambda_1-\lambda)f(\lambda_1-\lambda) f(\lambda_2-\lambda)}
-\frac{\l(\lambda_1-\lambda)\sigma(\lambda_1-\lambda_2)}{p(\lambda_1-\lambda) p(\lambda_1 -\lambda_2)}
 \right ]
\EN
and by using the relation
\EQ
\frac{\sigma(\lambda_1-\lambda) m(\lambda_2-\lambda) p(\lambda_1-\lambda_2)}{\sigma(\lambda_1-\lambda_2)} +
f(\lambda_1-\lambda)f(\lambda_2-\lambda) =  p(\lambda_1-\lambda)l(\lambda_2-\lambda)
\EN
we find that the contribution of $B(\lambda)$ is
\EQ
l(\lambda)^L \prod_{i=1}^{2} \frac{l(\lambda_i-\lambda)}{f(\lambda_i-\lambda)}
\EN

Finally, the operator $ \sum_{a} A_{aa}(\lambda) $ generates 
the terms which carry the hidden $6$-vertex structure . They have
been collected as
\bear
f(\lambda)^L l(\lambda_2)^L (F^{12}-F^{21}) -\frac{\sigma(\lambda_1-\lambda_2)}{p(\lambda_1-\lambda_2)}
\left \{ \frac{2 g(\lambda-\lambda_1)}{f(\lambda-\lambda_1)}[ 1- \frac{q^2(\lambda-\lambda_1}{g^2(\lambda-\lambda_1)}] +
\right . \nonumber \\ \left .
\frac{\sigma(\lambda-\lambda_2)
q(\lambda-\lambda_1) p(\lambda_1-\lambda_2)[r_{11}^{11}(\lambda-\lambda_1) +r_{21}^{21}(\lambda-\lambda_1)]}
{p(\lambda-\lambda_2)f(\lambda-\lambda_1) g(\lambda-\lambda_1) \sigma(\lambda_1-\lambda_2)}
\right . \nonumber \\ \left .
-\frac{\sigma(\lambda-\lambda_1) m(\lambda-\lambda_2) p(\lambda_1 -\lambda_2)}{f(\lambda-\lambda_1) 
p(\lambda -\lambda_1) f(\lambda-\lambda_2) \sigma(\lambda_1- \lambda_2)}
 \right \}
\ear

Remarkably enough the terms on the bracket of equation (B.30) can be factorized only in terms of the $6$-vertex $r$-matrix as
\EQ
\prod_{i=1}^{2} \frac{1}{f(\lambda-\lambda_i)} \left \{ r_{11}^{11}(\lambda-\lambda_1) r_{21}^{12}(\lambda-\lambda_2) 
+r_{12}^{21}(\lambda-\lambda_1) r_{22}^{22}(\lambda-\lambda_2) -r_{21}^{21}(\lambda-\lambda_1) r_{12}^{12}(\lambda -\lambda_2) 
\right \}
\EN
 
However, by using the auxiliary eigenvalue equation (61) for two sites, one is able to establish the following identity
\bear
\Lambda^{(1)}(\lambda,\{ \lambda_i \} )(F^{12}-F^{21}) =
 \left \{ r_{11}^{11}(\lambda-\lambda_1) r_{21}^{12}(\lambda-\lambda_2) 
+r_{12}^{21}(\lambda-\lambda_1) r_{22}^{22}(\lambda-\lambda_2) 
\right . \nonumber \\ \left .
-r_{21}^{21}(\lambda-\lambda_1) r_{12}^{12}(\lambda -\lambda_2) 
\right \} (F^{12}-F^{21})
\ear
and as a consequence of the expressions (B.31) and (B.32) 
we have that the eigenvalue contribution of $\sum_{a}A_{aa}(\lambda) $ is
\EQ
 f(\lambda)^L \prod_{i=1}^{2} \frac{\Lambda^{(1)}(\lambda,\{ \lambda_i \})}{f(\lambda - \lambda_i)} 
\EN

Hence, this completes the proof that the 
two-particle state (45) is the eigenstate of the $spl(2|1)$ model .
\vspace{0.5cm}
     
\centerline{\bf Appendix C : The three particle state }
\setcounter{equation}{0}
\renewcommand{\theequation}{C.\arabic{equation}}

This appendix is mainly concerned with the symmetrization property (52) of the three particle state . We begin our
discussion with the $ \lambda_2 \leftrightarrow \lambda_3 $ permutation . In this case, the three particle vector (see
equation (48) of section 5 ) becomes 
\bear
\vec{\Phi}_3(\lambda_1,\lambda_2,\lambda_3) = \vec{B}(\lambda_1) 
\otimes \vec{B}(\lambda_2) \otimes \vec{B}(\lambda_3) +
[l(\lambda_3)]^L \frac{\sigma(\lambda_2 -\lambda_3)}
{p(\lambda_2-\lambda_3)} \vec{B}(\lambda_1) \otimes F(\lambda_3) 
\vec{\xi}  
\nonumber \\
+ [l(\lambda_3)]^L \frac{\sigma(\lambda_1 -\lambda_3)}{p(\lambda_1-\lambda_3)} F(\lambda_1) \vec{\xi} \otimes \vec{B}(\lambda_2) 
\hat{F}_2(\lambda_3,\lambda_2)  
+[l(\lambda_2)]^L \frac{\sigma(\lambda_1 -\lambda_2)}{p(\lambda_1-\lambda_2)} F(\lambda_1) \vec{\xi} \otimes \vec{B}(\lambda_3) 
\hat{F}_3(\lambda_3,\lambda_2) \nonumber \\  
\ear

In order to relate this vector with the symmetric one $\vec{\Phi}_3(\lambda_1,\lambda_2,\lambda_3) $, we turn $\vec{B}(\lambda_2)$
over $\vec{B}(\lambda_3) $ with the help of the commutation rule (27) and as a result we get 
\bear
\vec{B}(\lambda_1) \otimes \vec{B}(\lambda_2) 
\otimes \vec{B}(\lambda_3) = \left \{ 
\vec{B}(\lambda_1) \otimes \vec{B}(\lambda_3) 
\otimes \vec{B}(\lambda_2) +
\right . \nonumber \\ \left .
\frac{\sigma(\lambda_3)-\lambda_2)}{p(\lambda_3-\lambda_2)} [l(\lambda_2)]^L 
\vec{B}(\lambda_1) \otimes F(\lambda_3) \vec{\xi}
\right \} . \frac{r_{23}(\lambda_2 -\lambda_3)}{
l(\lambda_2-\lambda_3)} \nonumber \\
-[l(\lambda_3)]^L \frac{\sigma(\lambda_2 -\lambda_3)}{p(\lambda_2-\lambda_3)} 
\vec{B}(\lambda_1) \otimes F(\lambda_2) \vec{\xi}  
\ear

The last term of this identity cancels out the 
second term in the vector $\vec{\Phi}_3(\lambda_1,\lambda_2,\lambda_3) $ and
the symmetrization rule 
\EQ
\vec{\Phi_3}(\lambda_1,\lambda_2,\lambda_3) = \vec{\Phi}_3(\lambda_1,\lambda_3,\lambda_2) . \frac{r_{23}(\lambda_2-\lambda_3)}{
l(\lambda_2 - \lambda_3) }
\EN
is valid provided the functions $\hat{F}_2(\lambda_2,\lambda_3) $ and 
$\hat{F}_3(\lambda_2,\lambda_3)$ satisfy the
following equations
\EQ
\hat{F}_3(\lambda_3,\lambda_2)= \hat{F}_2(\lambda_2,\lambda_3) l(\lambda_2-\lambda_3) r_{23}^{-1}(\lambda_2-\lambda_3)
\EN
\EQ
\hat{F}_2(\lambda_3,\lambda_2)= \hat{F}_3(\lambda_2,\lambda_3) l(\lambda_2-\lambda_3) r_{23}^{-1}(\lambda_2-\lambda_3)
\EN
but in fact they are equivalent  
since we have the following inversion properties
\EQ
l(x) l(-x) =1 ,~~ r_{23}(x) r_{23}(-x) =I_{23}
\EN

Similar reasoning can be implemented for the permutation $ \lambda_1 \leftrightarrow \lambda_2 $ . In this case however,
besides turning $\vec{B}(\lambda_1) $ over $ \vec{B}(\lambda_2) $, we also have to turn $\vec{B}(\lambda_1) $ over
$F(\lambda_2) \vec{\xi} $. By using the following commutations rule
\begin{displaymath}
F(\lambda) B_a(\mu) = \frac{q(\lambda-\mu)}{l(\lambda -\mu)} F(\mu) B_a(\lambda) + 
 \frac{g(\lambda-\mu)}{l(\lambda -\mu)} B_a(\mu) F(\lambda),~~a=1,2
\end{displaymath}
\EQ
B_a(\lambda) F(\mu) = \frac{q(\lambda-\mu)}{l(\lambda -\mu)} B_a(\mu) F(\lambda) + 
 \frac{g(\lambda-\mu)}{l(\lambda -\mu)} F(\mu) B_a(\lambda),~~a=1,2
\EN

This leads us with the term
\EQ
l(\lambda_2)^L \frac{\sigma(\lambda_1 -\lambda_2)}{p(\lambda_1-\lambda_2)} 
F(\lambda_1) \vec{\xi} \otimes \vec{B}(\lambda_3) 
\hat{F}_2(\lambda_2,\lambda_3)  
\EN
which appears on the vector $\vec{\Phi}_3(\lambda_1,\lambda_2,\lambda_3) $ 
and 
has to be canceled out by
\EQ
-l(\lambda_2)^L \frac{\sigma(\lambda_1 -\lambda_2)}{p(\lambda_1-\lambda_2)} F(\lambda_1) \vec{\xi} \otimes \vec{B}(\lambda_3) 
\frac{l(\lambda_3-\lambda_2)}{f(\lambda_3 -\lambda_2)}
\EN
and therefore we find that the function $\hat{F}_2(\lambda_2,\lambda_3) $ 
satisfies
\EQ
\hat{F}_2(\lambda_2,\lambda_3) = \frac{l(\lambda_3 -\lambda_2)}{f(\lambda_3 -\lambda_2)} I
\EN

Hence, equations (C.4) and (C.10) are able to fix  
the functions $\hat{F}_2(\lambda_2,\lambda_3)$ and 
$ \hat{F}_3(
\lambda_2,\lambda_3)$, and as a consequence follows the
expressions (49) and (50) of section 5. We have also checked 
the consistency of these results by verifying 
that all other terms satisfy the  condition of symmetry (51) of section 5
. Such calculation involves additional 
properties, since we need to turn twice
the operators $\vec{B}(\lambda_i)$ . Here we list some extra 
identities which are extremely useful to prove the symmetry under
$\lambda_1 \leftrightarrow \lambda_2 $, 
\EQ
[\vec{\xi} \otimes \vec{B}(y)]. r_{12}(x) = 
\frac{1-x}{1+x} \vec{\xi} \otimes \vec{B}(y), ~~
[\vec{B}(y) \otimes \vec{\xi}]. r_{23}(x) = \frac{1-x}{1+x} \vec{B}(y) \otimes \vec{\xi} 
\EN 
\EQ
[\vec{B}(y) \otimes \vec{\xi}]. r_{12}(x) =  \vec{B}(y) \otimes \vec{\xi} + r_{21}^{12}(x) \vec{\xi} \otimes \vec{B}(y),~~
[\vec{\xi} \otimes \vec{B}(y)]. r_{23}(x) =  \vec{\xi} \otimes \vec{B}(y) + r_{21}^{12}(x) \vec{B}(y) \otimes \vec{\xi}
\EN

Finally, another useful test is to look for certain unwanted terms which must be automatically canceled out. For instance, this
is the case of the terms
\EQ
[l(\lambda_2)]^L B_a(\lambda) E_a(\lambda_1) B_a(\lambda_3),~~ 
[l(\lambda_3)]^L E_a(\lambda) E_b(\lambda_1) B_c(\lambda_2) 
\EN

It is direct to verify that the first term in eq.(C.13) is canceled out with the help of $\hat{F}_2(\lambda_2,\lambda_3)$
while the second term depends on the functional form of the function $\hat{F}_3(\lambda_2,\lambda_3) $ .


\newpage
\centerline{ \bf Figures }
Figure 1. The 36 nonvanishing Boltzmann weights of the rational $spl(2|1)$ model.

\end{document}